\documentclass[11pt,english]{article}

\usepackage{mathptmx}

\usepackage[T1]{fontenc}
\usepackage[latin9]{inputenc}
\usepackage{babel}
\usepackage{float}
\usepackage{amstext}
\usepackage{geometry}
\usepackage{xcolor}
\usepackage{adjustbox}
\usepackage{pdflscape}

\usepackage[authoryear]{natbib}

\usepackage{amsmath, amssymb, graphicx, setspace, booktabs,
            tabularx, threeparttable, makecell, array, url, enumitem}
\usepackage{esint}

\usepackage[
  pdfusetitle,
  bookmarks=true,
  bookmarksnumbered=false,
  bookmarksopen=false,
  breaklinks=false,
  pdfborder={0 0 0},
  pdfborderstyle={},
  backref=false,
  colorlinks=false
]{hyperref}

\usepackage{chngcntr}
\usepackage{subcaption} 

\begin{document}

\begin{titlepage}
\thispagestyle{empty}

\title{\noindent Screening Out the Needy: \\The Effects of SNAP Work Requirements\thanks{ We thank Nathan Hendren and Stefanie Stantcheva for helpful comments. We also thank Nichole Szembrot for her support with data access and disclosure reviews and Matthew Unrath for his help with SNAP administrative data. Any views expressed are those of the authors and not those of the U.S. Census Bureau. The Census Bureau has reviewed this data product to ensure appropriate access, use, and disclosure avoidance protection of the confidential source data used to produce this product. This research was performed at the Cornell Federal Statistical Research Data Center (supported by the Cornell Center for Social Sciences) under FSRDC Project Number 3049 (CBDRB-FY26-P3049-R12788, CBDRB-FY26-P3049-R13048).
}}
\author{
Lexin Cai\thanks{E-mail: \protect\href{mailto:lexincai@g.ucla.edu}{lexincai@g.ucla.edu}} \\ UCLA
\and
Hyewon Kim\thanks{E-mail: \protect\href{mailto:hwkim@kdi.re.kr}{hwkim@kdi.re.kr}}\\ Korea Development Institute
\and 
Pauline Leung\thanks{E-mail: \protect\href{mailto:pleung@cornell.edu}{pleung@cornell.edu}}\\ Cornell University
}

\date{September 2026}
\maketitle

\begin{abstract}

\begin{singlespace}

{\normalsize
We examine the effectiveness of work requirements as a screening device in the Supplemental Nutrition Assistance Program (SNAP). Work requirements for ``able-bodied adults without dependents'' were suspended after the Great Recession and gradually reinstated across counties and states in the 2010s. Using linked administrative SNAP and employment data from five states and a triple-differences design, we find that work requirements reduce SNAP participation by seven percent without increasing labor supply and disproportionately screen out low-income individuals. We develop a welfare framework to interpret these results and find that the social costs of work requirements exceed budget savings.
}

\bigskip



\end{singlespace}

\end{abstract}

~
~

\thispagestyle{empty}

\end{titlepage}

\pagestyle{plain}
\setcounter{footnote}{0}

\section{Introduction\label{sec:Introduction}}

Work requirements have long been used to screen recipients of public assistance, on the logic that conditioning aid on work deters individuals with higher earnings capacity while preserving access for those most in need \citep{nichols1982targeting, besley1992workfare}. The premise remains influential today: work requirements now exist in cash, food, and medical assistance programs in the United States and are actively debated for housing assistance. The One Big Beautiful Bill Act, signed into law in July 2025, represents the largest expansion of work requirements for able-bodied adults on the Supplemental Nutrition Assistance Program (SNAP) since their introduction in 1996, and is projected to cause 2.4 million people to lose SNAP in an average month over the next decade \citep{cbo2025snap}.

In this paper, we examine the effectiveness of work requirements as a screening device in SNAP. ``Able-bodied adults without dependents'' (ABAWDs) on SNAP are typically subject to stringent work requirements; however, these rules were suspended during the Great Recession and gradually reinstated county-by-county throughout the 2010s. Using linked administrative SNAP and employment records from five states, we exploit this geographic variation together with the age-50 ABAWD eligibility cutoff in a triple-differences design. We find that reinstatement reduced SNAP participation by 4.4 percentage points, or seven percent relative to the baseline. Despite this sharp drop in caseloads, work requirements do not appear to have increased employment or earnings among affected individuals. Moreover, work requirements disproportionately screened out lower-income participants, worsening the targeting of benefits toward those with the greatest need. We do not find evidence that work requirements disproportionately screen out those who were not employed.


We then consider the welfare implications of these results. Work requirements are an example of an ``ordeal mechanism'' (\citealp{nichols1982targeting}), and \citet{finkelstein2019take} show that in a standard model, the targeting properties of ordeals are uninformative about social welfare. This is because in  a model where all parties are behaving optimally, participants who exit in response to the ordeal (the ``marginals'') are approximately indifferent between participating and not, so their welfare loss is negligible by the envelope theorem. The social cost of the ordeal is thus borne by ``inframarginal'' participants (who stay on the program) while the social benefits equal the fiscal savings from ``marginals'' exiting. Ordeals with poor targeting can therefore be welfare-improving as long as budget savings outweigh inframarginal hassle costs.

Yet, the public discourse and empirical literature on work requirements and other ordeals focus precisely on the harm to those pushed off the rolls (e.g., \citealp{herd2025administrative}). The literature has reconciled this disconnect by introducing behavioral biases:  \citet{finkelstein2019take}, for example, model non-participants as misperceiving program benefits, which restores a first-order role for exiting participants when assessing the welfare impacts of an ordeal-reducing intervention. 

We propose an alternative way of modeling work requirements that restores a first-order welfare role for those who exit, without appealing to behavioral biases. While standard ordeal models treat a policy change as a marginal increase in participation costs, work requirements may be effectively impossible for some participants to satisfy or document. We therefore allow compliance costs to vary across participants: for some, the requirement is merely a hassle, while for others it is infeasible. For this latter group, exit is not an optimizing choice at the margin, and their forgone benefits (net of saved participation costs) constitute a first-order social cost. The social costs of a work requirement are thus borne both by ``inframarginal'' participants (through compliance hassles) and by exiting participants (through lost benefits), and are weighed against the social benefits of saved program costs. Interestingly, although our approach is conceptually distinct from the misperception model of \citet{finkelstein2019take}, we show that under the calibration assumptions used in their paper, the two approaches deliver the same calibrated welfare calculations.

Calibrating our model, we find that accounting for the costs borne by exiting participants roughly doubles the social costs of work requirements across various parameterization assumptions: in our baseline specification, we estimate that the marginal value of public funds (MVPF; \citealp{hendren2020unified}) of work requirements, or the welfare costs to participants per dollar of budgetary savings, increases from 1.27 to 2.19 when exiters are included. When we incorporate social weights that reflect preferences for more redistribution, the difference is even more pronounced, precisely because we find that work requirements screen out lower-income participants. 

Our paper contributes to several strands of literature. First, it builds on a growing body of work estimating the effects of SNAP ABAWD work requirements on program participation and labor supply. These studies consistently find that work requirements reduce SNAP caseloads, while evidence on labor supply effects is mixed (\citealp{ribar2010food}; \citealp{ritter2018incentive}; \citealp{stacy2018impact};   \citealp{harris2021snap}; \citealp{wheaton2021impact9states}; \citealp{cuffey2022work}; \citealp{han2022impact}; \citealp{gray2023employed}). Our paper has three main advantages over this literature. First, our administrative records allow us to measure outcomes precisely and do not suffer from the underreporting of SNAP benefits that plagues survey data \citep{meyer2022errors}. Furthermore, they allow us to focus on recent SNAP participants, the group most directly affected by work requirements, rather than ``likely SNAP participants'' (based on income or education) commonly used in survey-based studies.  Second, relative to other studies using administrative data, we study additional states that allow for the use of a triple-differences design, which leverages both the age-50 eligibility cutoff and within-state geographic variation, rather than just one of these sources of variation. Third, our SNAP administrative data contain more detail on benefit levels and income, which allow us to study composition effects in addition to SNAP participation and labor supply outcomes. We compare estimates across studies alongside our results in Section \ref{sec:Results} and in Appendix Table \ref{tab:lit_compare_combined}.

Second, our paper contributes to a broader literature on work requirements across safety net programs and populations. Work requirements were a central feature of the 1996 welfare reform, which conditioned benefit receipt on work-related activities. A large literature examining the period leading up to welfare reform generally suggests that welfare-to-work programs reduced welfare receipt and increased employment, though the interpretation of these effects is complicated by the fact that work requirements were typically implemented alongside other policy changes, such as time limits and sanctions, while welfare reform itself coincided with Earned Income Tax Credit expansions as well as broader macroeconomic changes (see \citealp{blank2002evaluating,grogger2005welfare,ziliak2016temporary} for reviews). More recent work has revisited work requirements in contemporary safety net programs beyond the ABAWD work requirements we study here, including SNAP's general work requirements, Medicaid, and Temporary Assistance for Needy Families (TANF). Consistent with what we find for ABAWDs, \citet{cook2026disenrollment} find that SNAP's general work requirements for parents reduced participation without improving labor market outcomes. Evidence from Medicaid points in a similar direction: \citet{sommers2019medicaid} find that Arkansas's Medicaid work requirements reduced coverage during initial implementation but did not increase employment. Finally, for TANF, \citet{falk2023effects} finds that work requirements in Alabama increased employment but also the share of months families spent with neither earnings nor cash assistance, while \citet{richard2025penalties} find that increasing work sanction severity in Michigan reduced safety-net attachment and long-term formal employment, with short-term earnings gains too small to offset lost benefits.

Third, this paper contributes to the literature on how ordeals affect the targeting of public assistance. Ordeals---including transaction costs, stigma, and administrative burdens---can either improve or worsen targeting depending on whether they are more costly for high- or low-income individuals (\citealp{nichols1982targeting}; \citealp{deshpande2019screened}). We find that ABAWD work requirements disproportionately screen out lower-income participants and worsen targeting, consistent with evidence that some ordeals screen out recipients who are just as or more needy than average participants (\citealp{bhargava2015psychological}; \citealp{homonoff2021program}; \citealp{gray2023employed}; \citealp{shepard2025ordeals}), and in contrast to settings where ordeals improve self-selection (\citealp{alatas2016self}; \citealp{finkelstein2019take}; \citealp{anders2022welfare}; \citealp{rafkin2023selftargeting}; \citealp{mattunrath2024targeting}). In addition to this empirical contribution, we develop a welfare framework to interpret these targeting results. We show that worse targeting does not necessarily reduce welfare in a standard model (e.g., \citealp{finkelstein2019take}), and that allowing compliance to be infeasible for some participants  ties the targeting results back to social welfare. 


\section{Background and Institutional Context\label{sec:Background}}

Federal law imposes a work requirement on ABAWDs participating in SNAP, requiring at least 80 hours per month, often described as 20 hours per
week.\footnote{ABAWDs are also subject to general work requirements that apply to all SNAP recipients aged 16 to 59, including work registration, Employment and Training program participation (if assigned), and not voluntarily quitting a job.} ABAWDs can satisfy this requirement through any combination of paid or volunteer work, participation in SNAP Employment and Training (E\&T), or workfare programs (i.e., assigned community service).\footnote{Required workfare hours are generally calculated by dividing the participant's benefit by the state minimum wage.} Individuals who fail to meet the ABAWD work requirement are subject to a time limit and can receive SNAP benefits for only three months in a 36-month period. Some SNAP participants are exempt from ABAWD work requirements if they meet certain criteria, including having physical or mental barriers to work, or being pregnant. Moreover, states have a limited number of discretionary exemptions.\footnote{During our study period, states could exempt up to 15 percent of the ABAWD caseload subject to work requirements. State agencies have flexibility in how these are applied. For example, they may prioritize individuals who are working but unable to meet the 20-hour-per-week requirement, lack a high school diploma, or have limited English proficiency.}


States can request waivers for ABAWDs who reside in areas with weak labor market conditions, thus exempting them from work requirements. Geographic waivers are granted when the local unemployment rate is above 10 percent or at least 20 percent above the national average. States also qualify for a waiver when unemployment rates meet the criteria for  Extended Benefits (EB) under the Unemployment Insurance (UI) system. Waivers are typically approved for one year and can be applied at the level of states, counties, or sub-county areas. 

The scope of geographic waivers expanded significantly during the Great Recession. Under the American Recovery and Reinvestment Act (ARRA), SNAP work requirements were temporarily suspended nationwide from April 2009 through September 2010. Afterward, statewide waivers were broadly maintained until 2015 because many states' EB programs remained active. By 2016, the requirements were reinstated at least partially in over 40 states. Our paper focuses on the reinstatement of ABAWD work requirements for the first time since the Great Recession, which allows us to identify the effects for a  pool of recipients who participated in SNAP in the absence of ABAWD work requirements.

Table~\ref{tab:state-reinstatement} summarizes the timing of reinstatements across nine states in our sample. Our triple-differences analysis uses data from five states in which work requirements were partially reinstated, comparing counties where work requirements were reinstated against counties within the same state where waivers remained in place. Our difference-in-differences analysis uses data from all states, focusing only on counties that reinstated work requirements. Counties that changed waiver status during the 18-month follow-up period are excluded from both analyses.

\section{Data and Sample\label{sec:Data}}

\subsection{Data Sources}

We use individual-level administrative SNAP records from nine states: Colorado, Connecticut, Hawaii, Maryland, Massachusetts, Mississippi, New Jersey, North Carolina, and South Carolina. These consist of monthly observations containing demographic information, benefit amounts, gross income, and residential addresses. Gross income includes both earned (wages and self-employment) and unearned (e.g., social insurance, government transfers, and child support) income, and is only observed for months in which households are receiving SNAP benefits.\footnote{Earned and unearned income are not separately available for all states.} For most participants, income and benefits are unchanged within a recertification period (only changes in income are required to be reported), typically every 6 to 12 months.\footnote{Colorado, North Carolina, and South Carolina have a 6-month recertification period, while Connecticut and Massachusetts have a 12-month period; other states use a mix of recertification lengths.} While gross income and benefits are reported at the household level, we construct per-person measures by dividing them by household size. 


The administrative SNAP records are linked to employment and earnings from the Census's Longitudinal Employer-Household Dynamics (LEHD) records. Employment is defined as an indicator for having any UI-covered employment in the U.S. in a given quarter. Although we observe quarterly employment for all states in our sample, earnings data are available for only four states (Colorado, Connecticut, Maryland, and New Jersey). Earnings are measured as total quarterly earnings across all UI-covered jobs in the state of residence. If an individual has no employment or earnings record, the employment and earnings are coded as zero. All monetary amounts are expressed as 2016 dollars, using the Consumer Price Index (CPI-U). 

\subsection{Sample Construction}

Our analysis sample consists of individuals who received SNAP at any point between one and 18 months before reinstatement in their county, defining a baseline pool of recipients who received SNAP while work requirements were suspended.  We further restrict the sample to individuals residing in ``reinstated'' or ``control'' counties as of one month prior to reinstatement, where reinstated counties are those that implemented the ABAWD time limit for the first time since the Great Recession and maintained the requirement continuously for at least 18 consecutive months, while control counties are those in which ABAWD waivers remained continuously in effect over the same period.\footnote{County identifiers are obtained through the Census Bureau's Master Address File Auxiliary Reference File when available, and otherwise inferred using street addresses or ZIP codes (via SAS geocoding procedures). We exclude individuals whose county cannot be determined.} Individuals residing in counties that do not fall into either category are excluded from the analysis. We obtain each county's monthly waiver status from the U.S. Department of Agriculture.\footnote{Historical waiver records are available from the U.S. Department of Agriculture at \url{https://www.fns.usda.gov/snap/abawd/waivers}.} 

We focus on childless individuals aged 45 to 55. Within this group, we define ABAWDs as individuals aged 45 to 49 as of one month prior to reinstatement, and non-ABAWDs as individuals aged 50 to 55, who are exempt from ABAWD work requirements.\footnote{Information on ABAWD status is rarely recorded in SNAP administrative data, and not all states have information on disability status. We therefore proxy ABAWD status using age in months and the absence of children in the household. This definition likely overstates the ABAWD population and attenuates the effects we find.} This age restriction allows us to compare individuals who are similar in age. We follow individuals for up to 18 months after reinstatement and observe outcomes for up to 18 months prior to reinstatement to assess the plausibility of parallel pre-trends. In the two states with staggered reinstatement dates within a year, North Carolina and New Jersey, we keep observations from 18 months before the earliest reinstatement date through 18 months after the latest reinstatement date, yielding 24 months of post-period data for North Carolina and 25 months for New Jersey.\footnote{The 18-month post-reinstatement period allows us to account for variation in recertification lengths across states.} Lastly, we exclude individuals reporting monthly gross income above \$100,000. For Connecticut, due to incomplete data, we drop the last three months post-reinstatement.

Table~\ref{tab:summary_stats} presents the characteristics of individuals in our analysis sample, measured as of one month prior to  reinstatement (or the most recent pre-reinstatement month for those not enrolled at that point). ABAWDs (aged 45 to 49) and non-ABAWDs (50 to 55) differ on several dimensions: ABAWDs are younger and lower-income, but have higher employment rates and earnings. Across geographic areas, those with waivers have lower incomes and higher local area unemployment rates. However, any level differences and common shocks across age groups are differenced out by the triple-differences estimator. Demographic characteristics such as gender, race, and household size are similar across samples.

\section{Empirical Strategy\label{sec:Empirical-Strategy}}


\subsection{Triple-Differences Strategy}
Our main empirical strategy employs a triple-differences (DDD) design, comparing changes in outcomes across age groups and over time between counties that reinstated ABAWD work requirements and those in the same state where waivers remained in place. The identifying assumption for this DDD design is that, in the absence of reinstatement, the difference in outcomes between individuals aged 45 to 49 and those aged 50 to 55 would have evolved similarly in reinstated and control counties. This design differences out any age-specific trends that are common across counties. Because this specification relies on control counties, we implement it only for Colorado, Connecticut, Maryland, Massachusetts, and New Jersey, and exclude states where the work requirements were reinstated statewide (Hawaii, North Carolina, South Carolina, and Mississippi). We call this the ``partial states'' sample.

We estimate the following event-study specification separately for each state $s$ and then take a weighted average across states. The weights are equal to the state's share of treated individuals (i.e., aged 45 to 49 residing in treated counties). We report pooled estimates because our current data-sharing agreement does not allow for single-state analyses.  
\begin{align} 
\label{eq:Triple_ES}
Y_{iact} &= \sum_{k=-T, k \neq -1}^{T} \delta_{s,k} D_{it}^k + \omega_{ac} + \lambda_{at} + \gamma_{ct} + \epsilon_{iact} 
\end{align}
where $Y_{iact}$ denotes the outcome of interest for individual $i$ of age $a$ (prior to reinstatement), residing in county $c$ (prior to reinstatement),\footnote{Throughout, $c$ denotes a town in Connecticut and Massachusetts and a county in all other states.} and observed at time $t$. $D_{it}^k$ is a set of event-time indicators equal to 1 if individual $i$ is in the ``treated'' group (less than 50 years old and residing in a reinstated county as of one month prior to reinstatement) and $t$ is $k$ periods after the reinstatement month. We exclude $k = -1$, so all coefficients are relative to the month or quarter immediately before reinstatement. Individuals are observed for up to 18 months before and after reinstatement ($T=18$), except when outcomes are measured using LEHD data (e.g., employment), where periods are measured in quarters and $T=6$. The model includes age-by-county fixed effects $\omega_{ac}$, age-by-time fixed effects $\lambda_{at}$, and county-by-time fixed effects $\gamma_{ct}$. Standard errors are clustered at the county level. 

We estimate equation (\ref{eq:Triple_ES}) with SNAP participation (indicator for participation and benefit level) and labor supply (indicator for employment and earnings level) as outcomes. For composition effects, we estimate equation (\ref{eq:Triple_ES}) using only the sample of program participants (i.e., conditional on being on SNAP at time $t$), where the outcome is a fixed characteristic (e.g., gross income or benefit level measured right before reinstatement). 

To summarize the magnitudes, we estimate the following DDD specification. 
\begin{align}
\label{eq:Triple_DD}
Y_{iact} &= \beta_s \, Treat_{act} + \omega_{ac} + \lambda_{at} + \gamma_{ct} + \epsilon_{iact} 
\end{align}
where $Treat_{act}$ equals one in post-reinstatement periods for individuals aged 45 to 49 who resided in reinstated counties at baseline. For this specification, we drop the first three months following each state's reinstatement because ABAWDs are allowed a grace period of three months before they lose SNAP benefits. 

\subsection{Difference-in-Differences Strategy}
One disadvantage of the DDD design is that we can only implement it for states where there are both treated (``reinstated'') counties and counties where waivers remained in place. Therefore, we also use an alternative strategy that allows us to analyze all nine states: a difference-in-differences (DiD) design that restricts attention to counties where ABAWD work requirements were reinstated. We compare outcomes for individuals aged 45 to 49, who are subject to ABAWD work requirements, to individuals aged 50 to 55, who are exempt.  The disadvantage, relative to the DDD design, is that we need to rely on the more restrictive assumption that 50 to 55 year-olds are comparable to 45 to 49 year-old ABAWDs and that absent reinstatement, outcomes for these two age groups would have followed parallel trends.\footnote{In a robustness check, we restrict the sample to those age 48 to 52.} We report estimates for two samples: ``all states'', which includes all nine states in our sample, and ``partial states'', which enables comparison with DDD estimates.

As before, we estimate the event-study specification separately for each state and aggregate using a weighted average across states.
\begin{align}
\label{eq:Reinstate_ES}
Y_{iact} &= \sum_{k=-T, k \neq -1}^{T} \delta_{s,k} D_{it}^k + \alpha_a + \phi_t + \epsilon_{iact}
\end{align}
where $D_{it}^k$ is a set of event-time indicators equal to 1 if individual $i$ is treated (i.e., aged less than 50) and $t$ is $k$ periods after the reinstatement date,  $\alpha_a$ are age-in-month fixed effects, and $\phi_t$ are time fixed effects. $T$ is defined as above. Standard errors are clustered at the county level. We summarize the effects by estimating the DiD specification: 
\begin{align}
\label{eq:Reinstate_DD}
Y_{iact} &= \beta_{\text{s}} \, (ABAWD_a \times Post_t) + \alpha_a + \phi_t + \epsilon_{iact}
\end{align}
where $ABAWD_a$ is an indicator for being aged 45 to 49 and  $Post_t$ is an indicator for periods after reinstatement.

\section{Empirical Results\label{sec:Results}}

\subsection{Effects on Program Participation and Labor Supply}

\begin{figure}[!htbp]
  \caption{Effect of Work Requirements on SNAP Participation}
  \label{fig:event_study_participation}
  \centering
  
  \includegraphics[width=0.9\textwidth]{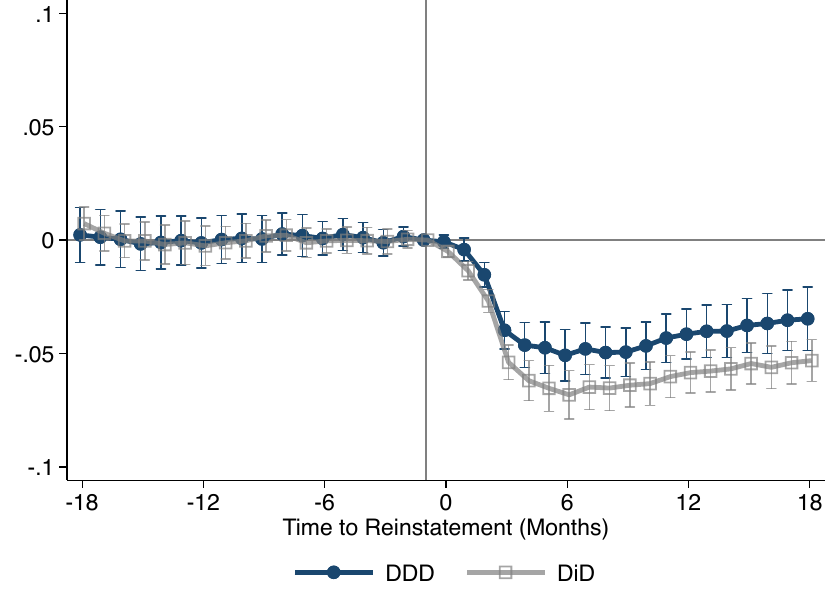}
  
  \vspace{0.5em}
  \begin{minipage}{0.9\textwidth}
    \footnotesize
    \textit{Notes:} This graph plots event-study estimates of Equations (\ref{eq:Triple_ES}) and (\ref{eq:Reinstate_ES}) where the outcome is monthly SNAP participation. The triple-differences (DDD) estimates are shown with solid dots, and the difference-in-differences (DiD) estimates are shown with square markers. Both the DDD and DiD samples include childless individuals aged 45 to 55 in five states that partially reinstated work requirements: Colorado, Connecticut, Maryland, Massachusetts, and New Jersey. The event-study specifications are estimated separately for each state and then aggregated using a weighted average across states. The weights are based on each state's share of individuals aged 45-49 residing in treated counties. Bars show 95\% confidence intervals.
  \end{minipage}
\end{figure}

We first examine how ABAWD work requirements affect SNAP participation. Figure~\ref{fig:event_study_participation} plots the coefficients from the DDD and DiD event-study specifications \eqref{eq:Triple_ES} and \eqref{eq:Reinstate_ES}. Both specifications show a parallel trend in the pre-period and a clear drop in participation three months after reinstatement of work requirements, though our preferred DDD specification shows a more muted effect. The decline in participation narrows slightly but a substantial gap persists through the end of the 18-month follow-up period. 

Table~\ref{tab:main_results} Panel A reports estimates of the decline in SNAP participation and benefits associated with Figure~\ref{fig:event_study_participation}. Column 1 shows the DDD estimates using equation \eqref{eq:Triple_DD} for our main ``partial states'' sample. We find that SNAP participation drops by 4.4 percentage points, which represents a seven percent reduction relative to the pre-reinstatement mean. Column 2 displays the DiD estimates using equation \eqref{eq:Reinstate_DD} for the same set of partial states. It shows a slightly larger magnitude---a 5.7 percentage-point reduction (9 percent). Finally, column 3 reports the DiD estimates using the set of all states, which shows an 8.7 percentage-point decline (13 percent) and indicates that the participation effects in states that reinstated work requirements statewide were larger than in states that only partially reinstated work requirements. Mirroring the participation results, the average SNAP benefits, including zeros for those who are no longer on the program, fall by \$8 (8 percent) in column 1; columns 2 and 3 show larger magnitudes.

Our results indicate variation across states in how much participation drops in response to work requirements. Cross-state heterogeneity may explain why we find a much smaller effect than the 37 percent drop for Virginia documented by \citet{gray2023employed} and the larger effects in most of the nine states studied by \citet{wheaton2021impact9states}. Although we cannot pinpoint the drivers of this heterogeneity, it is likely that institutional differences play a role. For example, high minimum wages may dampen the effects for two reasons: (i) participants are exempt from work requirements if they earn more than 30 times the federal minimum wage per week, which is easier to satisfy in high-wage states; and (ii) workfare hours needed to satisfy the requirement equal the participant's benefit divided by the state minimum wage.

\begin{figure}[!htbp]
  \caption{Effect of Work Requirements on Employment}
  \label{fig:event_study_emp_us}
  \centering
  
  \includegraphics[width=0.9\textwidth]{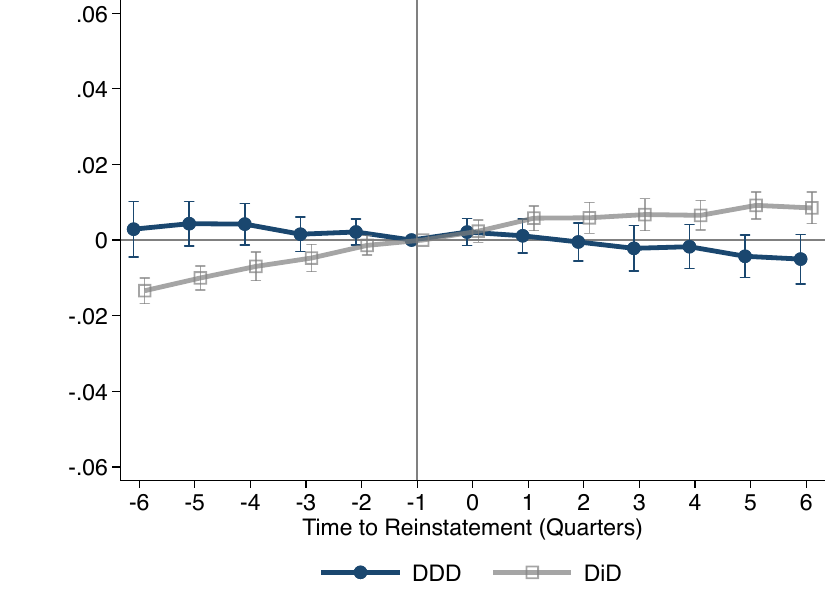}
  
  \vspace{0.5em}
  \begin{minipage}{0.9\textwidth}
    \footnotesize
    \textit{Notes:} This graph plots event-study estimates  of Equations (\ref{eq:Triple_ES}) and (\ref{eq:Reinstate_ES}) where the outcome is quarterly employment. The triple-differences (DDD) estimates are shown with solid dots, and the difference-in-differences (DiD) estimates are shown with square markers. Both the DDD and DiD samples include childless individuals aged 45 to 55 in five states that partially reinstated work requirements: Colorado, Connecticut, Maryland, Massachusetts, and New Jersey. The event-study specifications are estimated separately for each state and then aggregated using a weighted average across states. The weights are based on each state's share of individuals aged 45-49 residing in treated counties. Bars show 95\% confidence intervals.
  \end{minipage}
\end{figure}

Next, we turn to labor supply effects. Figure~\ref{fig:event_study_emp_us} shows the DDD and DiD event-study estimates from specifications \eqref{eq:Triple_ES} and \eqref{eq:Reinstate_ES}, where the outcome is an indicator for having UI-covered employment. When we use the DiD specification, the event-study graph displays a clear pre-trend prior to reinstatement, likely because the older comparison group has a different employment trend than ABAWDs. Indeed, when we use our DDD specification, which also compares across counties that did or did not reinstate work requirements, the pre-trends disappear and we find no statistically significant effects on employment. Given the event-study results, we rely on the DDD specification for employment effects. 

 Panel B of Table~\ref{tab:main_results}  shows that SNAP work requirements do not have a statistically significant effect on employment when we use the DDD specification (column 1). The point estimate is small, -0.4 percentage points, or 1 percent of the baseline mean. Our 95 percent confidence interval rules out effects beyond -1 percentage point and 0.2 percentage points. Columns 2 and 3 suggest positive effects; however, these estimates are likely biased given the pre-trends in Figure~\ref{fig:event_study_emp_us}.

When we examine how quarterly earnings (which are only available for Colorado, Connecticut, Maryland, and New Jersey) respond, we also find no statistically significant effects of work requirements, as shown in Appendix Figure~\ref{fig:event_study_earn} using the DDD specification. Column 1 of Table~\ref{tab:main_results} Panel B  shows a statistically insignificant reduction in quarterly earnings of \$15, or 1 percent of the baseline mean. Our 95 percent confidence interval rules out a decline greater than \$54 or an increase greater than \$23. Overall, we do not detect robust statistically significant effects on either employment or earnings.

\paragraph{Comparison with Existing Literature} Appendix Table~\ref{tab:lit_compare_combined} summarizes how our results compare with existing estimates.  Our participation effects are smaller than those in other administrative-data studies: \citet{gray2023employed} find a 37 percent drop in Virginia and \citet{wheaton2021impact9states} find effects of -45 to -18 percent across eight of nine states. Although our estimates are closer to studies that use nationally representative data \citep{stacy2018impact, harris2021snap,han2022impact}, it is worth noting that these studies typically focus on samples of ``likely SNAP participants'' that have lower baseline participation rates.
Our null employment effects are consistent with most studies using administrative data but have considerably tighter confidence intervals: our 95 percent confidence interval can rule out effects larger than four percent of the baseline, roughly four times narrower than \citet{gray2023employed}.\footnote{Although work requirements should in theory increase employment, \citet{wheaton2021impact9states} find \emph{negative} employment effects ranging from two to six percentage points with their main specification, though these effects disappear (or switch signs) in their sensitivity analysis; their largest positive effect is three percentage points in Missouri.} Some survey-based studies find positive employment effects (\citealp{harris2021snap}; \citealp{cuffey2022work}), but these differences likely reflect sample composition.\footnote{\citet{cuffey2022work} focus on low-income high school dropouts during 2005-2009, and \citet{harris2021snap} studies a sample with substantially higher baseline employment than ours.} Other survey-based studies (\citealp{ritter2018incentive}; \citealp{stacy2018impact}; \citealp{han2022impact}) find no labor supply effects.


\subsection{Composition Effects}

Next, we examine whether work requirements change the composition of SNAP recipients. We consider composition in terms of gross income, benefits, employment, gender, and race and ethnicity. To do this, we replace the outcomes in specifications \eqref{eq:Triple_ES} and \eqref{eq:Reinstate_ES} with a fixed characteristic  (e.g., income in the month prior to reinstatement of work requirements) \emph{conditional} on being a participant.\footnote{We can alternatively look at the ``dual'' of this measure by first defining subgroups based on fixed characteristics and examining the drop in participation for each subgroup (e.g., \citealp{deshpande2019screened}, who use both approaches). We prefer our approach for uncluttered visualization as we only need to plot one line per outcome.} Therefore, these results show how the composition of the SNAP caseload changes as a result of work requirements. 

\begin{landscape}
\begin{figure}[!htbp]
  \centering
  \caption{Composition Effects}
  \label{fig:composition_fixed}

  \begin{subfigure}{0.6\textwidth}
    \centering
    \caption{Gross Income}
    \includegraphics[width=\textwidth]{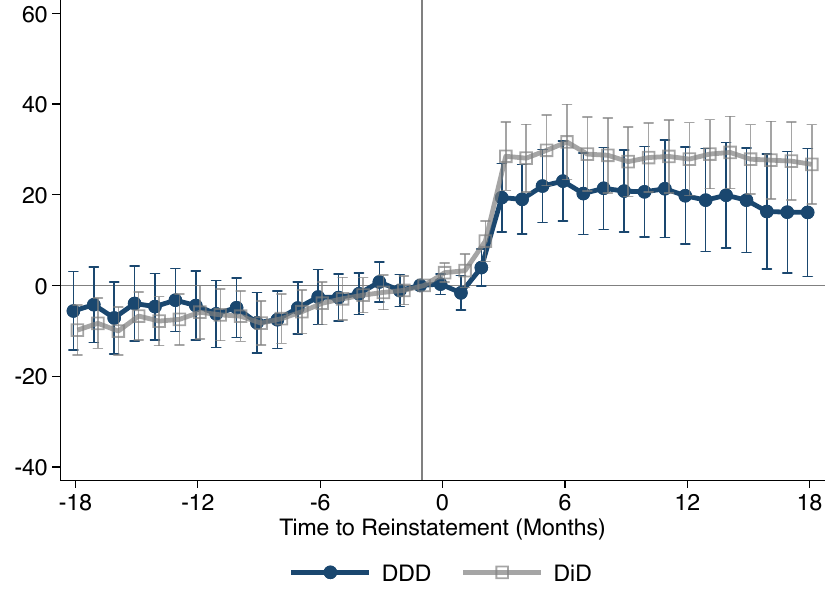}
    \label{fig:event_study_gross_inc_per_person_fixed}
  \end{subfigure}
  \hfill
  \begin{subfigure}{0.6\textwidth}
    \centering

    \caption{SNAP Benefits}
    \includegraphics[width=\textwidth]{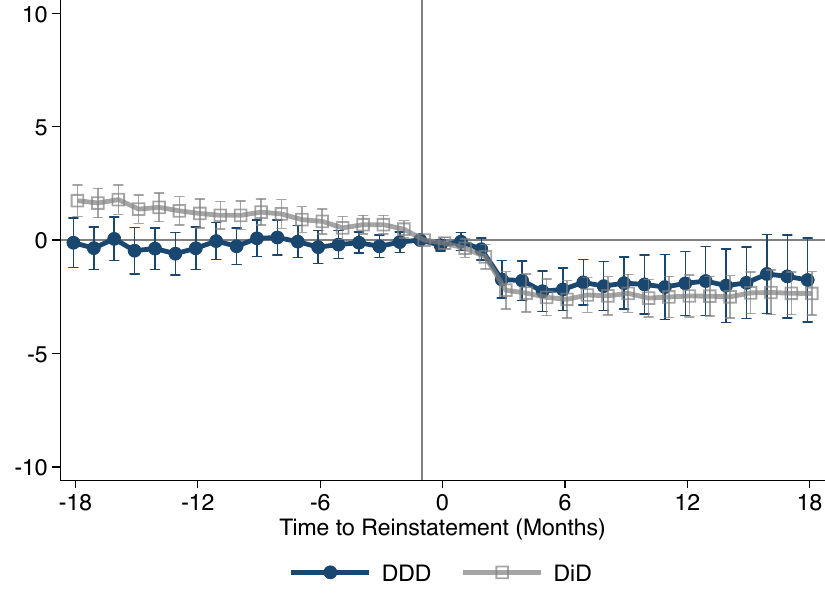}
    \label{fig:event_study_benefit_per_person_fixed}

  \end{subfigure}

  \vspace{-1.5em}

  \begin{subfigure}{0.6\textwidth}
    \centering

     \caption{Receiving Maximum SNAP Benefit}
    \includegraphics[width=\textwidth]{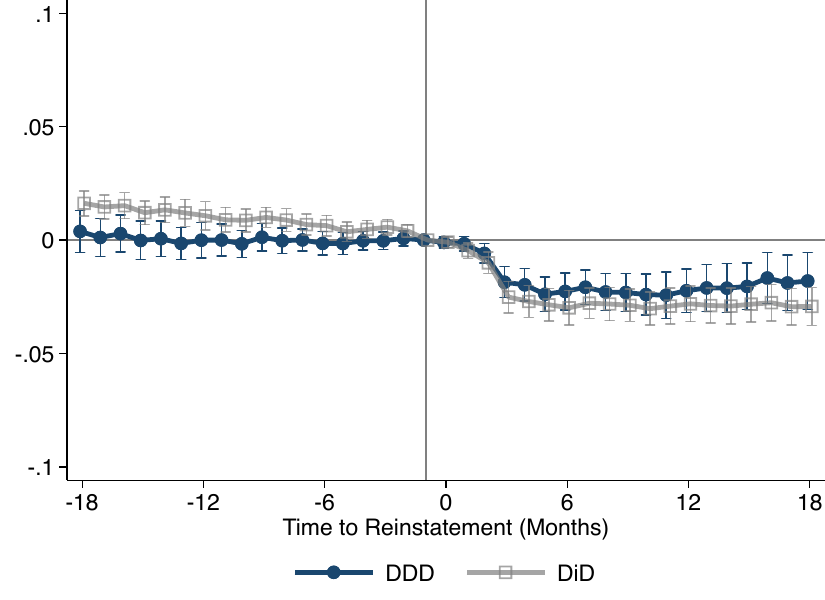}
    \label{fig:event_study_maxben_fixed}
   
  \end{subfigure}
  \hfill
  \begin{subfigure}{0.6\textwidth}
    \centering
    \caption{Employed}
    \includegraphics[width=\textwidth]{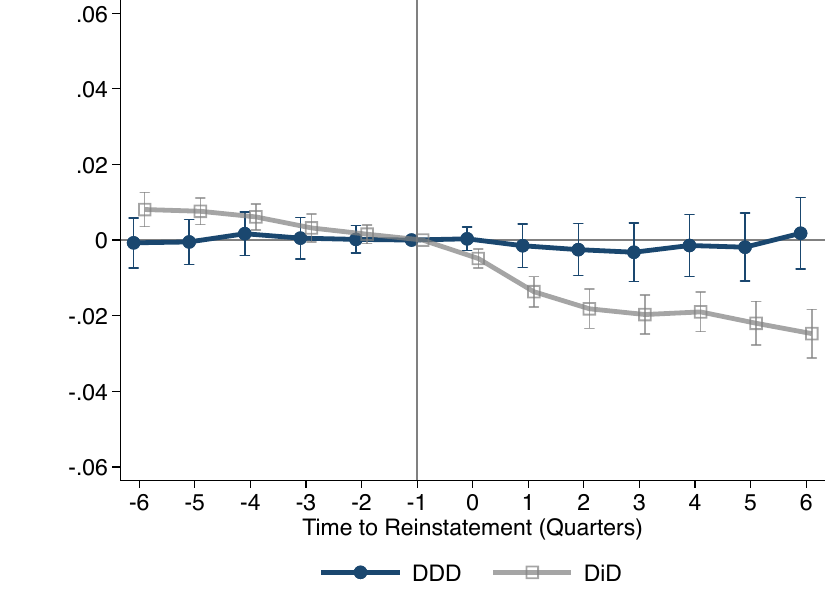}
    \label{fig:event_study_emp_us_1yr}
  \end{subfigure}

  \vspace{-1.35em}

  \begin{minipage}{1.4\textwidth}
    \footnotesize
    \textit{Notes:} This graph plots event-study estimates of Equations (\ref{eq:Triple_ES}) and (\ref{eq:Reinstate_ES}) where the outcomes are: monthly gross income per person, SNAP benefits per person, an indicator for receiving maximum benefits,  and an indicator for employment in any of the last four quarters, all measured in the month or quarter prior to reinstatement of work requirements. The sample contains only individuals participating in SNAP at each event time. The triple-differences (DDD) estimates are shown with solid dots, and the difference-in-differences (DiD) estimates are shown with square markers. Both the DDD and DiD samples include childless individuals aged 45 to 55 in five states that partially reinstated work requirements: Colorado, Connecticut, Maryland, Massachusetts, and New Jersey. The event-study specifications are estimated separately for each state and then aggregated using a weighted average across states. The weights are based on each state's share of individuals aged 45-49 residing in treated counties. Bars show 95\% confidence intervals.
  \end{minipage}

\end{figure}

\end{landscape}

Figure~\ref{fig:composition_fixed} (and accompanying Table \ref{tab:composition_fixed}) shows that work requirements reduced participation for relatively low-income individuals more than higher-income individuals. The average gross income of participants (measured pre-reinstatement) rises by \$24 (4 percent) after reinstatement (panel a), while the average benefit and the share receiving the maximum benefit (both measured pre-reinstatement) fall by \$2 (1 percent) and 2 percentage points (5 percent), respectively (panels b and c).\footnote{In Appendix Figure~\ref{fig:not_fixed}, we show versions of these graphs that do not hold incomes (benefits) fixed at their pre-reinstatement values, but reflect the incomes (benefits) of participants at the event time. These figures therefore reflect both a composition effect and potentially a behavioral response among the inframarginal participants (i.e., those who stay). We find that the effects on gross income and receiving the maximum benefit are 63 and 50 percent larger, respectively, when we do not fix the incomes (benefits), but most of the estimated change at the reinstatement of work requirements appears to be coming from compositional changes. We also report the associated DiD and DDD estimates in Appendix Table~\ref{tab:composition_notfixed}.}

The final panel of Figure~\ref{fig:composition_fixed}  examines how the proportion of participants who are employed changes as a result of work requirements. A stated reason for imposing work requirements is to incentivize those who are not employed to either work or disenroll from the program. We find no evidence of this: We do not detect a statistically significant change in the proportion of employed individuals in the caseload, with employment defined as having been employed in any of the four quarters prior to reinstatement, using our main DDD specification. 

Appendix Table~\ref{tab:demographics} analyzes compositional changes by gender, race, and ethnicity by reporting DDD and DiD estimates for these caseload characteristics. We find a statistically significant 0.5 percentage-point increase (1 percent) in the proportion of (non-Hispanic) White, but do not detect robust statistically significant changes in the female share or in other race/ethnicity categories.

Overall, we find that work requirements disproportionately disenrolled people with lower incomes. This is consistent with some of the findings of \citet{gray2023employed}, who report that those who are homeless and without earned income experience the largest drops in participation.\footnote{\citet{gray2023employed} also find that those with a disability exemption experience a smaller drop. They do not have SNAP benefit or income amount information.} In Section \ref{sec:welfare_framework}, we examine the welfare consequences of composition changes in the SNAP caseload.

\subsection{Robustness to Alternative Specifications and Samples}

We conduct two analyses to assess the robustness of our results to alternative specifications and sample definitions. First, we account for staggered treatment timing using the approach of \citet{sun2021estimating}. This applies to North Carolina and New Jersey, where policy changes occurred at different points within a year. Our specifications for New Jersey (DiD and DDD) and for North Carolina (DiD) are linear two-way fixed effects (TWFE) regressions, which may create problematic ``negative weights'' for some treatment cohorts when treatment effects are heterogeneous \citep{de2020two}. Although such bias is likely limited in our context given that we have large never-treated groups and closely staggered event times, we implement a cohort-specific event-study specification.\footnote{Although similar TWFE-related problems can arise when using conventional triple-difference regressions in staggered settings, \citet{ortiz2025better} and \citet{strezhnev2023decomposing} show that using never-treated units as controls avoids this bias. The vast majority of our control group consists of never-treated units; only our New Jersey DDD (TWFE) estimates contain some not-yet-treated controls. The cohort-interacted specification in this section helps address this concern by estimating cohort-specific event-time effects relative to never-treated controls, rather than pooling not-yet-treated cohorts as controls.} Specifically for these two states, we modify equations \eqref{eq:Triple_ES} and \eqref{eq:Reinstate_ES} by interacting the treatment indicator with cohort-specific event-time indicators and then aggregate cohort-specific estimates (\citealp{sun2021estimating}). Table~\ref{tab:adj_staggering} reports the results, which are similar to Table~\ref{tab:main_results}. 

Second, to address the possibility that 45 to 49 year-olds are not comparable to 50 to 55 year-olds, we reconstruct our sample using a tighter age range of 48 to 52, as in  \citet{han2022impact}. A disadvantage of this comparison is that a larger proportion of the ``treated'' 48 to 49 year-olds will eventually age out of work requirements during the follow-up period. We report the results in Table~\ref{tab:ages48_to_52}. As expected due to the aging out of work requirements, the participation magnitudes are more muted than those in Table~\ref{tab:main_results}, though overall patterns remain similar.

\section{Welfare Effects of Work Requirements\label{sec:welfare_framework}}

\subsection{Model}

We now assess the social welfare consequences of a work requirement
by adapting the framework of \citet{finkelstein2019take}.

There are two types, $j\in\{H,L\}$, of individuals who earn wages $\theta_{j}$
in the labor market, choose hours of work $h_{j}$, and pay taxes
$\tau(\theta_{j}h_{j})$. A government assistance program pays benefit
$B_{j}$ but imposes a utility cost of $\Lambda\kappa_{j}+c$, where
$c$ is an individual-specific participation cost and $\kappa_{j}$
is a type-specific cost that can be changed by policy parameter $\Lambda$.\footnote{This is an application cost in \citet{finkelstein2019take}, but we
will assume that if someone applies, they will participate.} Individuals have utility 
\begin{align*}
u(y_{j}^{P}+B_{j})-v(h_{j}^{P})-(\Lambda\kappa_{j}+c) & \quad\text{if participating,}\\
u(y_{j}^{\neg P})-v(h_{j}^{\neg P}) & \quad\text{otherwise,}
\end{align*}
where $y_{j}^{P}=\theta_{j}h_{j}^{P}-\tau(\theta_{j}h_{j}^{P})$ and
$y_{j}^{\neg P}=\theta_{j}h_{j}^{\neg P}-\tau(\theta_{j}h_{j}^{\neg P})$
are post-tax labor income for participants and non-participants of
type $j$. Individuals participate when utility from doing so exceeds
the outside option, which defines a threshold 
\[
c_{j}^{*}=u(y_{j}^{P}+B_{j})-v(h_{j}^{P})-u(y_{j}^{\neg P})+v(h_{j}^{\neg P})-\Lambda\kappa_{j}
\]
 where only those with $c<c_{j}^{*}$ participate.

Since we find no labor supply responses, we model the work requirement purely as a compliance cost rather than a mandated change in hours. A standard way to model this cost is to
assume that the ordeal parameter increases from $\Lambda$ to $\bar{\Lambda}$
(as in \citealp{finkelstein2019take}). However, as we show below,
this implies that all those who exit the program are close to indifferent
between participating and not, and are therefore negligible in the
welfare calculation---a result that is at odds with public discourse.
An alternative is to assume that the ability to comply with work requirements varies across
individuals. For a fraction $1-q_{j}$ of type-$j$ participants,
the requirement is feasible to satisfy (e.g., they already work sufficient
hours, qualify for an exemption, or can complete workfare or training)
and merely shifts the ordeal parameter, as in the standard case. Utility
for feasible participants under the work requirement is 
\[
u(y_{j}^{P}+B_{j})-v(h_{j}^{P})-(\bar{\Lambda}\kappa_{j}+c)\qquad\text{if participating,}
\]
which lowers their participation threshold to $\bar{c}_{j}^{*}=c_{j}^{*}-(\bar{\Lambda}-\Lambda)\kappa_{j}$.
For the remaining fraction $q_{j}$, compliance is infeasible because they cannot secure sufficient documented work hours, a workfare assignment,
or an exemption determination.\footnote{We assume that feasibility is independent of $c$ within each type.}
For these individuals, a work requirement represents a loss of access to the program (rather than a choice). 

The social planner maximizes 
\begin{align*}
W=\sum_{j\in\{L,H\}}g_{j}\pi_{j}\Bigg[ & \int_{0}^{c_{j}^{*}}\bigl(u(y_{j}^{P}+B_{j})-v(h_{j}^{P})-(\Lambda\kappa_{j}+c)\bigr)\,dF_{j}(c)+\int_{c_{j}^{*}}^{\infty}\bigl(u(y_{j}^{\neg P})-v(h_{j}^{\neg P})\bigr)\,dF_{j}(c)\Bigg]
\end{align*}
subject to the budget constraint 
\[
\pi_{H}A_{H}B_{H}+\pi_{L}A_{L}B_{L}=\pi_{H}A_{H}G_{H}^{P}+\pi_{H}\left[1-A_{H}\right]G_{H}^{\neg P}+\pi_{L}A_{L}G_{L}^{P}+\pi_{L}\left[1-A_{L}\right]G_{L}^{\neg P}
\]
where $A_{j}=F_{j}(c_{j}^{*})$ is the participation rate of type
$j$, and $G_{j}^{P}=\tau(\theta_{j}h_{j}^{P})$ and $G_{j}^{\neg P}=\tau(\theta_{j}h_{j}^{\neg P})$
are the non-benefit portions of the government budget for participants
and non-participants. In the basic model, $G$ consists only of tax
revenue, but one could extend it to include, for example, administrative
costs associated with running the assistance program. $g_{j}$ are
social welfare weights and $\pi_{j}$ are population shares of each
type. Assuming as in \citet{finkelstein2019take} that the government
budget costs are borne by an individual with average marginal utility
of consumption in the population, the social welfare impact of work
requirements is 
\begin{eqnarray}
\Delta W & = & \underbrace{\sum_{j\in\{L,H\}}g_{j}\pi_{j}A_{j}^{1}\left(-\bar{\Lambda}+\Lambda\right)\kappa_{j}}_{\text{Mechanical effect on inframarginal participants' utility}}\nonumber \\
 &  & +\underbrace{\sum_{j\in\{L,H\}}g_{j}\pi_{j}(1-q_{j}) \int_{\bar{c}_{j}^{*}}^{c_{j}^{*}}\left[u(y_{j}^{\neg P})-v(h_{j}^{\neg P})-u(y_{j}^{P}+B_{j})+v(h_{j}^{P})+(\Lambda\kappa_{j}+c)\right]dF_{j}(c) }_{\text{Effect on feasible participants who exit }}\nonumber \\
 &  & +\underbrace{\sum_{j\in\{L,H\}}g_{j}\pi_{j}q_{j} \int_{0}^{c_{j}^{*}}\Big[u(y_{j}^{\neg P})-v(h_{j}^{\neg P})-u(y_{j}^{P}+B_{j})+v(h_{j}^{P})+(\Lambda\kappa_{j}+c)\Big]\,dF_{j}(c) }_{\text{Effect on participants for whom compliance is infeasible}}\nonumber\\
 &  & -\underbrace{\{ \pi_{H}\Delta A_{H}\left[G_{H}^{\neg P}-G_{H}^{ P}\right]+\pi_{L}\Delta A_{L}\left[G_{L}^{\neg P}-G_{L}^{ P}\right]\}}_{\text{Tax revenue effect from exiting participants}}\nonumber \\
 &  & -\underbrace{\left[ \pi_{H}\Delta A_{H}B_{H}+\pi_{L}\Delta A_{L}B_{L} \right]}_{\text{Program cost effect from exiting participants}}\label{eq:welfare_std}
\end{eqnarray}
where $A_{j}^{1}=(1-q_{j})F_{j}(\bar{c}_{j}^{*})$ and $A_{j}^{0}=F_{j}(c_{j}^{*})$
are participation rates under work requirements and no requirements,
respectively; $\Delta A_{j}=A_{j}^{1}-A_{j}^{0}$ is the change in
program participation.

When all participants face the same
marginal increase in the ordeal ($q_{j}=0$), the model collapses to the standard case: the third line of (\ref{eq:welfare_std})
is zero and the second line is negligible by the envelope theorem, since exiting participants are approximately indifferent
between participating and not. The welfare calculation then weighs
the hassle costs of the work requirements borne by those who stay on the program
(the ``inframarginals'') against the budgetary impacts; those who
exit do not enter the welfare calculation. When compliance is instead infeasible for some participants ($q_{j}>0$), the third
line remains, and for this group lost benefits have a first-order welfare cost. The second line remains negligible, and we set it to zero in what follows.

Equation (\ref{eq:welfare_std}) simplifies once we assume that there
are no labor supply responses ($h_{j}^{P}=h_{j}^{\neg P}$ for $j\in\{L,H\}$)
and apply first-order approximations to express the utility changes in money-metric terms (as in \citet{finkelstein2019take}): 
\begin{eqnarray}
\Delta W & \approx & -\sum_{j\in\{L,H\}}\tilde{g}_{j}\pi_{j}A_{j}^{1}\tilde{\kappa}_{j}(\bar{\Lambda}-\Lambda)\nonumber \\
 &  & +\sum_{j\in \{L,H\} }\tilde{g}_{j}\pi_{j}(-B_{j}+\Lambda\tilde{\kappa}_{j}+\bar{c}_{j})\,q_{j}A_{j}^{0}\nonumber \\
 &  & -\pi_{H}\Delta A_{H}B_{H}-\pi_{L}\Delta A_{L}B_{L}\label{eq:cal_standard}
\end{eqnarray}
where each utility term is normalized by that type's own marginal utility:
$\tilde{g}_{j}=g_{j}u'(y_{j}^{P}+B_{j})$, $\tilde{\kappa}_{j}=\frac{\kappa_{j}}{u'(y_{j}^{P}+B_{j})}$,
and $\bar{c}_{j}=\frac{1}{u'(y_{j}^{P}+B_{j})}\frac{\int_{0}^{c_{j}^{*}}cdF_{j}(c)}{\int_{0}^{c_{j}^{*}}dF_{j}(c)}$
for $j\in\{L,H\}$.

Expression (\ref{eq:cal_standard}) clarifies
the targeting implications. Consider the extreme case where the planner
cares only about $L$-types ($\tilde g_{H}=0$), the work requirement causes
all $L$-types to exit ($A_{L}^{1}=0$), and all $H$-types are unaffected
($\Delta A_{H}=0$). In a standard model where $q_{j}=0$, the welfare
impact would be $\Delta W\approx-\pi_{L}\Delta A_{L}B_{L}>0$ because
the government saves on benefits that would have gone to the disenrolled,
even though all targeted individuals exit. In a model where $q_{j}>0$,
the welfare impact is approximately 
\[
\Delta W\approx\pi_{L}\left[\tilde{g}_{L}-1\right]\Delta A_{L}B_{L}-\tilde{g}_{L}\pi_{L}(\Lambda\tilde{\kappa}_{L}+\bar{c}_L)\Delta A_{L}
\]
where we have used the approximation that $\Delta A_{L}$$\approx-q_{L}A_{L}^{0}$.\footnote{This assumes there are few true ``marginals'' ($F_j(c_{j}^{*})-F_j(\bar{c}_{j}^{*})\approx0$).}
The first term is negative whenever the social weight on $L$-types'
consumption exceeds the average taxpayer's ($\tilde{g}_{L}>1$). The
second term reflects the value of saved participation costs for those
who exit. Poor targeting now implies a welfare cost through the
first term. 

\citet{finkelstein2019take} show that an alternative way of bringing
exiting participants back into the welfare calculation is to assume
that agents misperceive benefits. Interestingly, we show in Appendix
\ref{sec:RelationshipFN} that under their misperception model and
calibration assumptions, \citet{finkelstein2019take} arrive at a
welfare expression identical to our calibrated equation (\ref{eq:cal_standard}).

\subsection{Calibration}

We follow \citet{finkelstein2019take} and summarize the welfare effects in equation (\ref{eq:cal_standard}) as a ratio of the participants' willingness to pay to the budgetary savings, or a ``marginal value of public funds'' (MVPF; \citealp{hendren2020unified}):\footnote{Note that the MVPF assumes $\tilde g_L=\tilde g_H=1$ (i.e., equal social weights across types and the average taxpayer), in keeping with the definition in \cite{hendren2020unified}. We report social-welfare-weighted versions of the MVPF below.}
\begin{equation}
\label{eq:mvpf}
 MVPF = \frac{\sum_{j} \pi_j A^1_j \tilde\kappa_j(\bar\Lambda-\Lambda)
      + \sum_{j} \pi_j (B_j - \Lambda\tilde\kappa_j - \bar c_j)\, q_j A^0_j}
     {-\pi_H \Delta A_H B_H - \pi_L \Delta A_L B_L}
\end{equation}
The numerator is the aggregate willingness to pay to avoid work requirements, while the denominator is the budget savings. An MVPF above one indicates that the welfare cost to participants exceeds the budget savings. Since a work requirement is a revenue-raising policy rather than a spending policy, a lower MVPF is more desirable \citep{boning2025welfare}. 

To parameterize the model, we define ``types'' similarly to \citet{finkelstein2019take}:
Type-$L$ are those who receive the maximum benefit (lowest income)
and type-$H$ are those who receive less than the maximum benefit.
In our data, $B_{L}=192$ and $B_{H}=93$. We estimate that the change
in participation for each type is $\Delta A_{L}=-0.065$ and $\Delta A_{H}=-0.022$
(and again use the approximation that $\Delta A_{j}\approx-q_{j}A_{j}^{0}$).\footnote{The exact expression is $\Delta A_{j}=-q_j A_j^0-(1-q_j)[F_j (c_j^* )-F_j (\bar c_j ^* )]$, so we are effectively assuming that all of the exiting participants are not marginal. If, however, a share $s$ of the exiters are actually marginal, the cost associated with the exiters will be roughly scaled by $1-s$ (because the welfare loss for true marginals is relatively small).}
We also observe each type's participation rate to be $A_{L}^{1}=0.59$
and $A_{H}^{1}=0.66$ with population shares $\pi_{L}=0.49$ and $\pi_{H}=0.51$
(in our sample of recent participants). We estimate the budgetary savings of work requirements---the denominator of (\ref{eq:mvpf})---to be $\$7.16$ per person per month.

For ordeal costs, we assume work requirements increase compliance
costs by two hours per month among inframarginal participants, valued
at the federal minimum wage of \$7.25 per hour in our baseline case,
yielding $\tilde{\kappa}_{j}(\bar{\Lambda}-\Lambda)=14.50$.\footnote{We were unable to find a study that estimates the time costs of complying
with work requirements, though it is likely highly variable depending
on whether someone is already stably employed and meeting the 80 hours
per month requirement (and therefore simply needs to provide documentation
during certification), exempt for other reasons, or not exempt and needs to fulfill the requirement via
job search or workfare.} Given the uncertainty in this number, we also present results for
compliance costs between 1 and 4 hours. For exiting participants'
average participation cost $\Lambda\tilde{\kappa}_{j}+\bar{c}_{j}$,
we follow \citet{finkelstein2019take}, who estimate an application (time) cost of \$75, which we assume is spread over a six-month recertification period (\$12.50
per month).

The first two columns of Table~\ref{tab:welfare_calcs_table} present the
MVPF when we exclude and include the costs to exiting participants, respectively, under various assumptions. At the baseline assumption of two hours of compliance time for inframarginals,
the MVPF is $1.27$ when we only account for hassle costs for inframarginal
participants (column 1), but increases to $2.19$ when we incorporate costs to the exiting participants (column 2). The rest of the table shows weighted MVPFs, which incorporate social weights $\tilde g_j$.\footnote{Specifically, we report $\bar g MVPF$ where $\bar g=\sum_j \tilde g_j\frac{WTP_j}{\sum _jWTP_j}$ and $WTP_j=\pi_j A^1_j \tilde\kappa_j(\bar\Lambda-\Lambda)+ \pi_j (B_j - \Lambda\tilde\kappa_j - \bar c_j)\, q_j A^0_j$.} When $\tilde{g}_{H}=0$ and $\tilde{g}_{L}=2$ (i.e., we do not
care about the $H$-types but we care twice as much about the $L$-types
as the average taxpayer), the weighted MVPFs are $1.17$ and $2.77$ excluding and including exiting participants, respectively. The
final columns assume that we value $L$-types twice as much as $H$-types
and average taxpayers, which increases the weighted MVPFs to $1.85$
and $3.58$. The rows show how our estimates change when we assume
different time costs for the inframarginal participants (i.e., $\tilde{\kappa}_{j}(\bar{\Lambda}-\Lambda)$).

In general, we find that accounting for the exiting participants
nearly doubles the welfare cost. This is especially true when the
planner has redistributive preferences because low-income participants
are disproportionately screened out. The quantitative importance of exiting participants is consistent with \citet{deshpande2019screened}, who estimate in the context of disability programs that lost benefits to deserving applicants deterred by an office closing can exceed budgetary savings by an order of magnitude.
\section{Conclusion\label{sec:Conclusion}}

Work requirements have been a cornerstone of social safety net policy for decades, premised on the dual goals of incentivizing employment and targeting benefits toward the neediest population. This paper evaluates whether SNAP's ABAWD work requirements achieve either of these objectives by leveraging linked administrative data and quasi-experimental variation in the reinstatement of requirements across counties following the Great Recession.

We find that work requirements reduce program participation by seven percent but generate no detectable improvement in employment or earnings. Importantly, we find that the participation losses are concentrated among lower-income recipients. In this sense, work requirements function as an ordeal that worsens, rather than improves, the targeting of the program toward those with the greatest need.

We develop a welfare framework to interpret these findings. The standard approach to evaluating ordeals models them as marginal increases in participation costs, so that, following the logic of the envelope theorem, participants who exit the program contribute negligibly to the welfare calculation. Under this framework, poor targeting is not necessarily socially costly. We argue that this framing fits poorly both with the nature of work requirements and with the public discourse around them, which centers on the harm to individuals pushed off the rolls. We instead model compliance as infeasible for a subset of participants, for whom the requirement eliminates access to the program. For this group, lost benefits are a first-order welfare cost. Under our calibrated model, we estimate that the welfare cost to participants of work requirements is $\$2.19$ per dollar of budget savings.

Our findings are relevant to the recent expansion of ABAWD work requirements under the One Big Beautiful Bill Act, which raises the ABAWD age ceiling to 64 and restricts waivers for weak local labor markets. These changes impose work requirements on an older, lower-employment population and in areas where qualifying work is hardest to find, which our results suggest is likely to reduce access to SNAP without generating meaningful employment gains.

One area for future work is understanding the sources of heterogeneity in the effects of work requirements. Our estimates vary across the states in our sample, and prior work has documented even larger participation effects in settings such as Virginia (\citealp{gray2023employed}). Understanding what drives the variation in effects would be valuable for designing policies that better balance the goals of targeting and supporting employment.

\pagebreak{}

\subsection*{Declaration of Gen-AI Technologies in the Writing Process}

During the preparation of this work, we used Claude 5 and ChatGPT 5.6 to streamline the paper and improve clarity. All substantive content is our own, and we have reviewed and edited all AI-assisted output. We take full responsibility for this article.

\pagebreak{}

\begin{singlespace}
\end{singlespace}

\begin{singlespace}
\bibliographystyle{aea}
\bibliography{references.bib}
\end{singlespace}

\pagebreak{}


\clearpage


\clearpage

\clearpage

\begin{table}[!htbp]
\centering
\footnotesize
\begin{threeparttable}
\caption{Timing and Coverage of ABAWD Work Requirement Reinstatements Across States}
\label{tab:state-reinstatement}
\setlength{\tabcolsep}{5pt}
\renewcommand{\arraystretch}{1.2}
\begin{tabularx}{\textwidth}{
  >{\raggedright\arraybackslash}p{2.5cm}
  >{\raggedright\arraybackslash}p{3.2cm}
  >{\raggedright\arraybackslash}X
  >{\raggedright\arraybackslash}X
  >{\raggedright\arraybackslash}X}
\toprule
State & Reinstatement dates & Reinstated areas & Control areas & Areas excluded from analysis \\
\midrule
\addlinespace[6pt]
\multicolumn{5}{l}{\textbf{Partial reinstatements}} \\
\addlinespace[6pt]
Colorado & 1/1/2016 & 33 counties & 13 counties & 18 counties \\
\addlinespace[3pt]
Connecticut & 1/1/2016 & 46 towns & 82 towns & 41 towns \\
\addlinespace[3pt]
Maryland & 1/1/2016 & 11 counties & 11 counties & 2 counties \\
\addlinespace[3pt]
Massachusetts & 1/1/2016 & 212 towns & 110 towns & 29 towns \\
\addlinespace[8pt]
New Jersey & \makecell[l]{1/1/2016 \\ 2/1/2016 \\ 5/1/2016 \\ 8/1/2016} &
\makecell[l]{4 counties \\ 3 counties \\ 7 counties \\ 4 counties} &
3 counties & 0 counties \\
\addlinespace[8pt]
\multicolumn{5}{l}{\textbf{Statewide reinstatements}} \\
\addlinespace[6pt]
Hawaii & 12/1/2014 & 5 counties & 0 counties & 0 counties \\
\addlinespace[3pt]
Mississippi & 1/1/2016 & 82 counties & 0 counties & 0 counties \\
\addlinespace[8pt]
North Carolina & \makecell[l]{1/1/2016 \\ 7/1/2016} &
\makecell[l]{23 counties \\ 77 counties} &
0 counties & 0 counties \\
\addlinespace[8pt]
South Carolina & 4/1/2016 & 46 counties & 0 counties & 0 counties \\
\bottomrule
\end{tabularx}

\vspace{1.0em}
\textit{Notes:} This table shows the dates and areas in which states reinstated work requirements for able-bodied adults without dependents (ABAWDs), based on data from \href{https://www.fns.usda.gov/snap/abawd/waivers}{the U.S. Department of Agriculture}. Reinstated areas are defined as those that implemented ABAWD work requirements for the first time since the Great Recession and maintained them continuously for at least 18 consecutive months, while control areas are those in which ABAWD waivers remained continuously in effect over the same period. Counties and towns that do not fall into either category are excluded from the analysis. See Appendix~\ref{sec:TimingReinstatements} for state-specific details.
\end{threeparttable}
\end{table}


\begin{landscape}
\begin{table}[t]
\caption{Summary Statistics}
\label{tab:summary_stats}
\vspace{-60bp}
\centering
\includegraphics[page=1,width=1.5\textwidth]{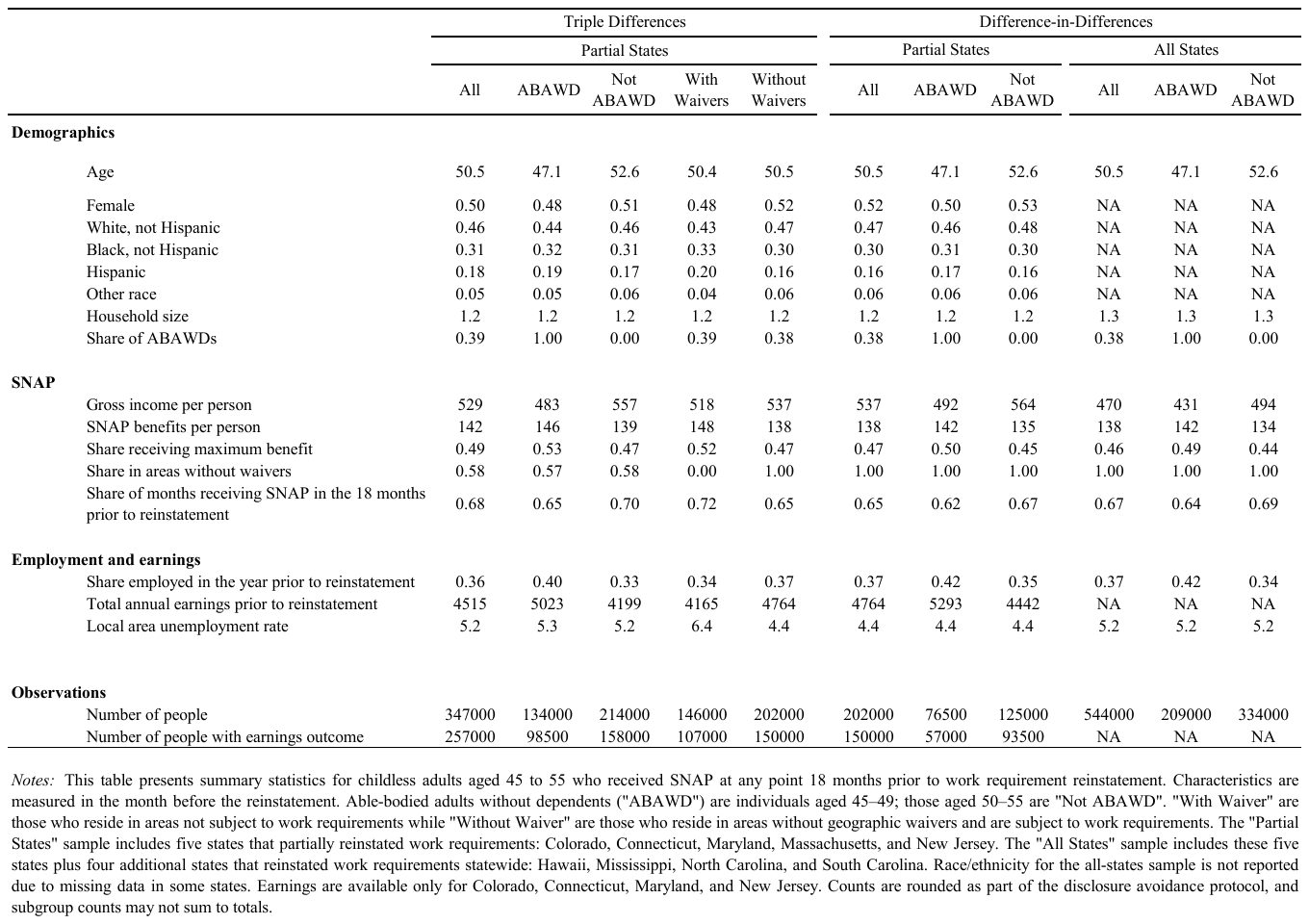}
\begin{centering}
\begin{minipage}[t]{1\columnwidth}%
\end{minipage}
\par\end{centering}
\end{table}
\end{landscape}


\begin{table}[t]
\caption{Effects of SNAP Work Requirements on Program Participation and Labor Supply}
\label{tab:main_results}
\vspace{-30bp}
\makebox[\textwidth][c]{\hspace*{2.35cm}\includegraphics[page=1,width=1.2\textwidth]{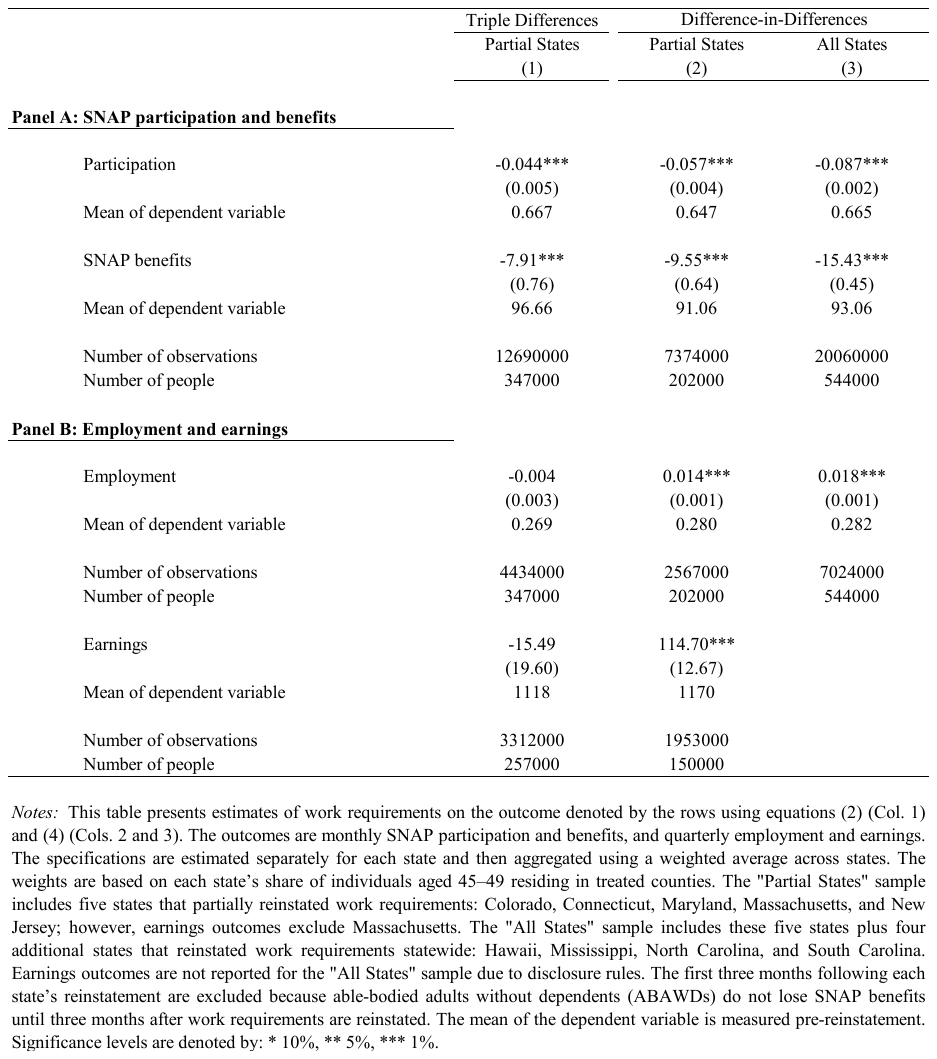}}
\begin{centering}
\begin{minipage}[t]{0.82\columnwidth}%
\end{minipage}
\par\end{centering}
\end{table}


\begin{table}[t]
\caption{Effects of SNAP Work Requirements on Characteristics of Program Participants}
\label{tab:composition_fixed}
\vspace{-30bp}
\makebox[\textwidth][c]{\hspace*{2cm}\includegraphics[page=1,width=1.2\textwidth]{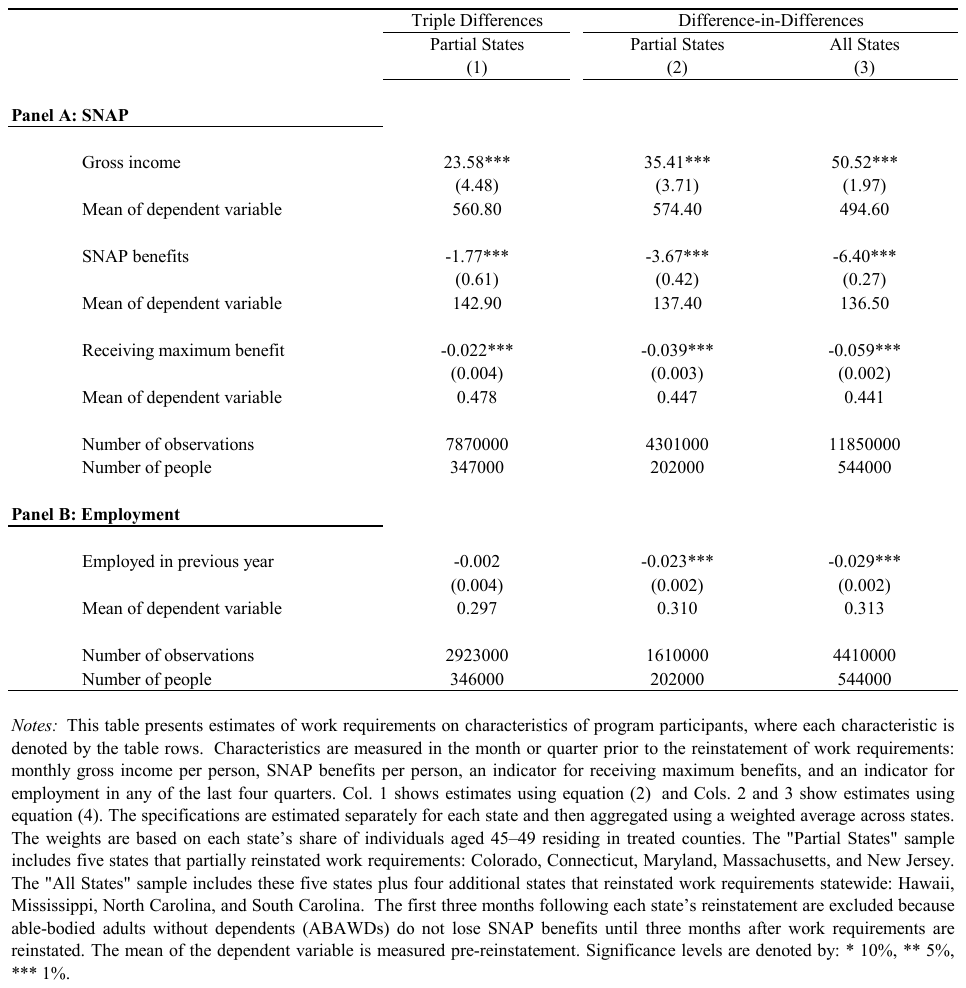}}
\begin{centering}
\begin{minipage}[t]{0.82\columnwidth}%
\end{minipage}
\par\end{centering}
\end{table}


\begin{table}[t]
\caption{Marginal Value of Public Funds Estimates}
\label{tab:welfare_calcs_table}
\vspace{-60bp}
\includegraphics[page=1,width=1.2\textwidth]{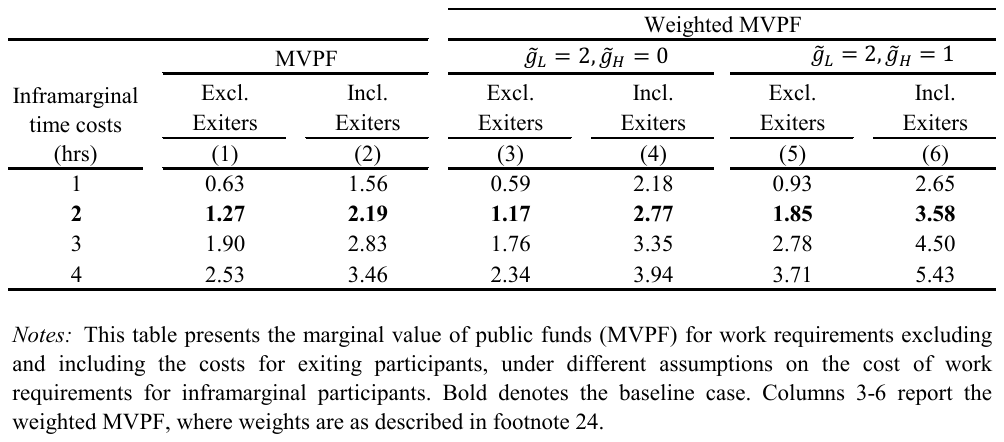}
\begin{centering}
\begin{minipage}[t]{0.82\columnwidth}%
\end{minipage}
\par\end{centering}
\end{table}

\clearpage
\appendix

\counterwithin{table}{section} 
\counterwithin{figure}{section}

\section*{Appendix}

\section{Relationship to \citet{finkelstein2019take}}

\label{sec:RelationshipFN}

In this section, we show how our model delivers the same calibrated
welfare expression as one using the misperception model of \citet{finkelstein2019take}.

Consider the model in Section \ref{sec:welfare_framework}. \citet{finkelstein2019take}
assume that $B_{H}=B_{L}=B$, $g_{H}=g_{L}=1$, and a unit mass of
each type; we suppress $\pi_{j}$ terms in what follows. They consider
an infinitesimally small reform that increases the ordeal $\Lambda$.\footnote{That is, unlike work requirements which also mandate minimum work
hours, the policy changes only the ordeal parameter. } Assuming that participants misperceive the benefit
to be $(1+\varepsilon_{j})B$ instead of $B$, they show that the social welfare
impact is 
\begin{eqnarray*}
\frac{dW}{d\Lambda} & = & -\kappa_{H}A_{H}-\kappa_{L}A_{L}\\
 &  & +\mu_{H}\frac{dA_{H}}{d\Lambda}+\mu_{L}\frac{dA_{L}}{d\Lambda}\\
 &  & +\frac{dA_{H}}{d\Lambda}\left[\tau(\theta_{H}h_{H}^{P})-\tau(\theta_{H}h_{H}^{\neg P})\right]+\frac{dA_{L}}{d\Lambda}\left[\tau(\theta_{L}h_{L}^{P})-\tau(\theta_{L}h_{L}^{\neg P})\right]-\left(\frac{dA_{H}}{d\Lambda}+\frac{dA_{L}}{d\Lambda}\right)B
\end{eqnarray*}
where $A_{j}=F_{j}(c_{j}^{*})$ is the proportion of each type $j$
that are program participants and $\mu_{j}=u(y_{j}^{P}+B)-u(y_{j}^{P}+(1+\varepsilon_{j})B)$
is the degree of misperception in utility terms.

For calibration, they assume $\varepsilon_{j}$ to be such that marginal
individuals are indifferent between applying for benefits and not
applying given misperceptions and the (time) cost of applying. Combined
with the fact that they do not find labor supply effects so that $h_{j}^{P}=h_{j}^{\neg P}$,
this means:\footnote{In their calibration exercise, \citet{finkelstein2019take} approximate
the utility terms with first-order Taylor expansions. The expression
becomes $(1+\varepsilon_{j})B-(\Lambda\tilde{\kappa}_{j}+\tilde{c}_{j}^{*})=0$,
where the $\tilde{\kappa_{j}}$ and $\tilde{c_{j}}$ are cost terms
that are expressed in dollars. They assume that time costs $\Lambda\tilde{\kappa}_{j}+\tilde{c}_{j}^{*}=\$75$
and infer $\varepsilon_{j}$.} 
\[
u\left(y_{j}^{P}+(1+\varepsilon_{j})B\right)-(\Lambda\kappa_{j}+c_{j}^{*})=u(y_{j}^{\neg P})
\]
Substituting this condition into $\mu_{j}$, the expression that they
calibrate for the welfare impact of a change in the ordeal parameter
is: 
\begin{eqnarray}
\frac{dW}{d\Lambda} & = & -\kappa_{H}A_{H}-\kappa_{L}A_{L}\nonumber \\
 &  & +\left[u(y_{H}^{P}+B)-u(y_{H}^{\neg P})-(\Lambda\kappa_{H}+c_{H}^{*})\right]\frac{dA_{H}}{d\Lambda}+\left[u(y_{L}^{P}+B)-u(y_{L}^{\neg P})-(\Lambda\kappa_{L}+c_{L}^{*})\right]\frac{dA_{L}}{d\Lambda}\nonumber \\
 &  & -\left(\frac{dA_{H}}{d\Lambda}+\frac{dA_{L}}{d\Lambda}\right)B\label{eq:FN}
\end{eqnarray}
The first line is the welfare impact on the inframarginal participants,
the second line is the impact on marginal participants, and the third
is the budget impact.

Now consider the welfare impact of a work requirement that is infeasible
to meet for some participants, again assuming that $B_{H}=B_{L}=B$,
$g_{H}=g_{L}=1$, and that there are no labor supply effects. Equation (\ref{eq:welfare_std})
in Section \ref{sec:welfare_framework} simplifies to 
\begin{eqnarray*}
\Delta W & = & \sum_{j\in\{H,L\}}A_{j}^{1}\left(\Lambda-\bar{\Lambda}\right)\kappa_{j}\\
 &  & -\sum_{j\in\{H,L\}}\left[u(y_{j}^{\neg P})-u(y_{j}^{P}+B)+(\Lambda\kappa_{j}+\bar{c}_{j})\right]\Delta A_{j}\\
 &  & -\left(\Delta A_{H}+\Delta A_{L}\right)B
\end{eqnarray*}
where $\bar{c}_{j}=\frac{\int_{0}^{c_{j}^{*}}cdF_{j}(c)}{\int_{0}^{c_{j}^{*}}dF_{j}(c)}$. 

The two expressions coincide term by term once we use the approximation $\Delta A_{L}$$\approx-q_{L}A_{L}^{0}$, and the derivative in (\ref{eq:FN}) is scaled by the size of the reform, 
$\frac{dA_{j}}{d\Lambda}(\bar\Lambda-\Lambda)\approx\Delta A_{j}$ and $-\kappa_{j}A_{j}(\bar\Lambda -\Lambda)\approx A_{j}^{1}(\Lambda-\bar{\Lambda})\kappa_{j}$. The one remaining difference is that the misperception model evaluates exiters'
participation costs at the marginal participant's cutoff, $\Lambda\kappa_{j}+c_{j}^{*}$,
while the infeasibility model evaluates them at the participant average,
$\Lambda\kappa_{j}+\bar{c}_{j}$. If we calibrate both with the same
value---\citet{finkelstein2019take} estimate participation costs
using time costs of typical applicants---the calibrated welfare calculations coincide numerically, even though the mechanism through which exiting participants contribute to social welfare is different.

\section{Timing of Reinstatement of Work Requirements Across States}
\label{sec:TimingReinstatements}

We summarize ABAWD work requirement implementation details and sample construction decisions for each state in our sample.

\subsection*{Colorado}
Colorado reinstated ABAWD work requirements in selected counties on January 1, 2016. Some counties had already administered ABAWD work requirements prior to 2016, coupled with mandatory E\&T programs. We exclude these previously treated counties from our analysis, along with counties that changed waiver status during the 18-month follow-up period.

\subsection*{Connecticut}
Connecticut reinstated ABAWD work requirements in selected towns/cities on January 1, 2016. Connecticut had waivers for both towns and counties; we classify towns and cities located within waived counties as waived. Some Connecticut towns/cities are excluded from the sample due to geographic identifier issues, along with towns and cities that changed waiver status during the 18-month follow-up period.

\subsection*{Maryland}
Maryland reinstated ABAWD work requirements in selected counties on January 1, 2016. We exclude counties that changed waiver status during the 18-month follow-up period.  

\subsection*{Massachusetts}
Massachusetts reinstated ABAWD work requirements in selected towns/cities on January 1, 2016. Massachusetts had waivers for both towns and counties; we classify towns and cities located within waived counties as waived. Some Massachusetts towns/cities are excluded from the sample due to geographic identifier issues, along with towns and cities that changed waiver status during the 18-month follow-up period.

\subsection*{New Jersey}
New Jersey had staggered implementation in January, February, May, and August 2016. Atlantic, Cape~May, and Cumberland Counties had the limits in effect only briefly before regaining waivers on January~1,~2017. Based on publicly available reports from the New Jersey Department of Human Services, caseload reductions were minimal in these counties (Atlantic: 71; Cape~May: 1; Cumberland: 45 people). We treat these counties as controls.

\subsection*{Hawaii}

Hawaii reinstated ABAWD work requirements on December 1, 2014. Although Molokai retained a waiver after reinstatement, the island has a population of around 7{,}000. Therefore we treat Hawaii as implementing the time limit statewide. 

\subsection*{Mississippi}
Mississippi reinstated ABAWD work requirements statewide on January 1, 2016. 

\subsection*{North Carolina}
North Carolina had staggered implementation in January and July 2016. North Carolina had a partial waiver covering 77 counties from January~1 to June~30, 2016, and then implemented the time limit statewide.

\subsection*{South Carolina}
South Carolina was approved for a statewide waiver from January through December~2016 but discontinued the waiver at the end of March~2016, effectively reinstating ABAWD work requirements statewide on April 1, 2016.


\clearpage 
\setcounter{figure}{0}
\renewcommand{\thefigure}{A.\arabic{figure}}

\begin{figure}[!htbp]
  \caption{Effect of Work Requirements on Earnings}
  \label{fig:event_study_earn}
  \centering
  
  \includegraphics[width=0.9\textwidth]{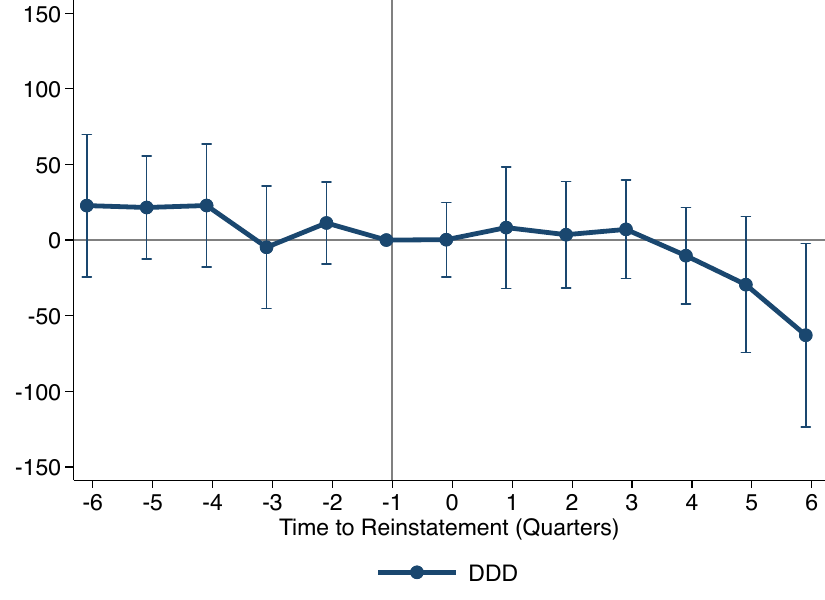}
  
  \vspace{0.5em}
  \begin{minipage}{1\textwidth}
    \footnotesize
    \textit{Notes:} This graph plots event-study estimates of Equation (\ref{eq:Triple_ES}) where the outcome is quarterly earnings (including zeros for those without earnings). The sample includes childless individuals aged 45 to 55 in four states that partially reinstated work requirements and have LEHD earnings: Colorado, Connecticut, Maryland, and New Jersey. The event-study specifications are estimated separately for each state and then aggregated using a weighted average across states. The weights are based on each state's share of individuals aged 45-49 residing in treated counties. Bars show 95\% confidence intervals.
  \end{minipage}
\end{figure}


\clearpage

\begin{landscape}
\begin{figure}[!htbp]
  \centering
  \caption{Composition Effects (Contemporaneous Characteristics)}
  \label{fig:not_fixed}

  \begin{subfigure}{0.6\textwidth}
    \centering

    \caption{Gross Income}
    \includegraphics[width=\textwidth]{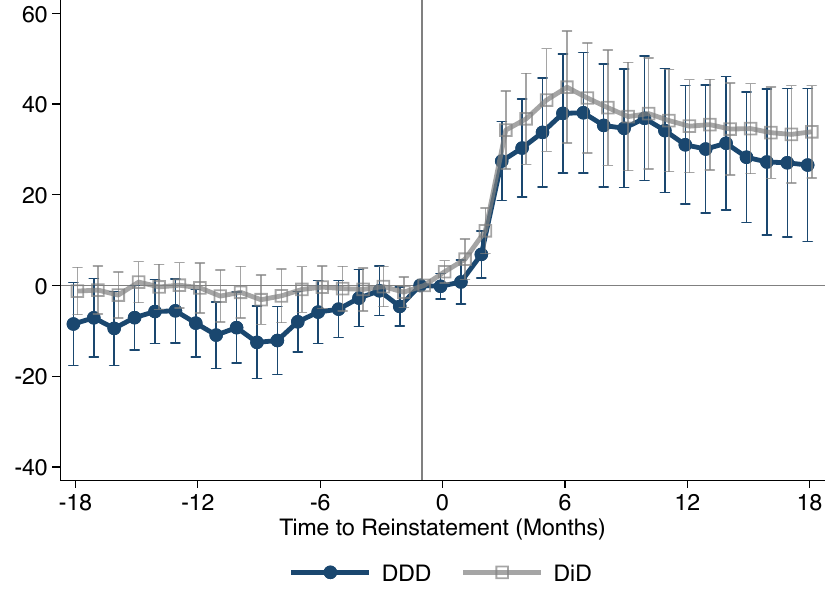}
    \label{fig:event_study_gross_inc_per_person}
  \end{subfigure}
  \hfill
  \begin{subfigure}{0.6\textwidth}
    \centering

     \caption{SNAP Benefits}
    \includegraphics[width=\textwidth]{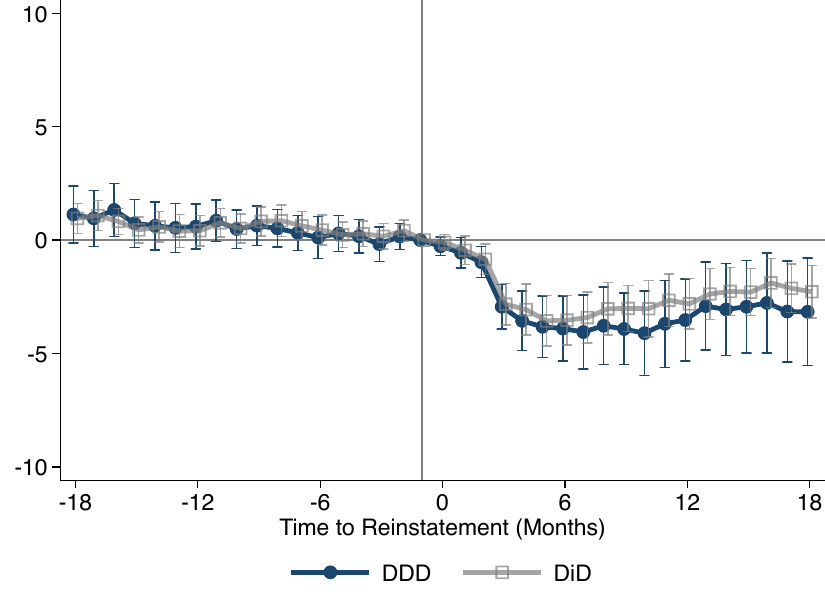}
    \label{fig:event_study_benefit_per_person}

  \end{subfigure}

  \vspace{-1.5em}

  \begin{subfigure}{0.6\textwidth}
    \centering

    \caption{Receiving Maximum SNAP Benefit}
    \includegraphics[width=\textwidth]{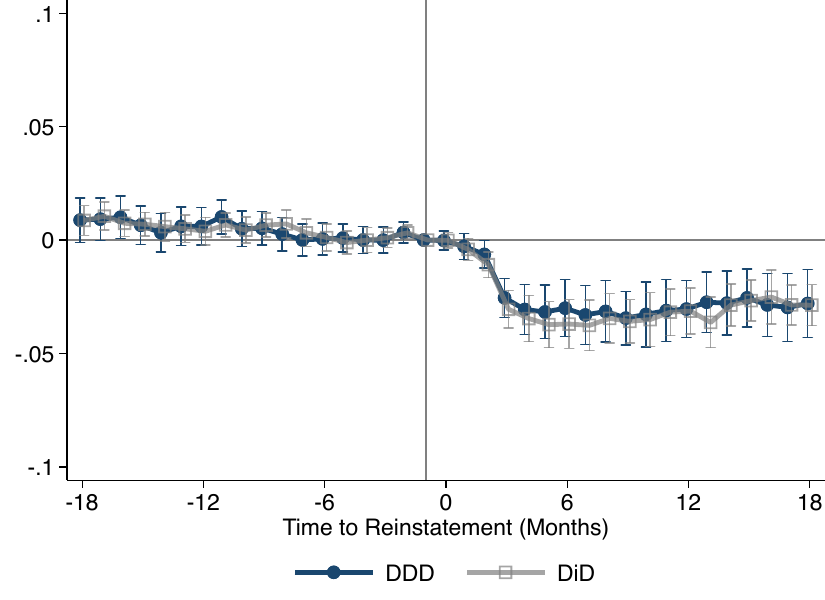}
    \label{fig:event_study_maxben_notfixed}
   
  \end{subfigure}
 
  \vspace{-1.35em}

  \begin{minipage}{1.4\textwidth}
    \footnotesize
    \textit{Notes:} This graph plots event-study estimates of Equations (\ref{eq:Triple_ES}) and (\ref{eq:Reinstate_ES}) where the outcomes are: monthly gross income per person, SNAP benefits per person, and an indicator for receiving maximum benefits. The sample contains only individuals participating in SNAP at each event time. The triple-differences (DDD) estimates are shown with solid dots, and the difference-in-differences (DiD) estimates are shown with square markers. Both the DDD and DiD samples include childless individuals aged 45 to 55 in five states that partially reinstated work requirements: Colorado, Connecticut, Maryland, Massachusetts, and New Jersey. The event-study specifications are estimated separately for each state and then aggregated using a weighted average across states. The weights are based on each state's share of individuals aged 45-49 residing in treated counties. Bars show 95\% confidence intervals.
  \end{minipage}

\end{figure}

\end{landscape}







\setcounter{table}{0}
\renewcommand{\thetable}{A.\arabic{table}}


\begin{landscape}
\begin{table}[htbp]\centering
\caption{Comparison of Estimates from Prior Literature}
\label{tab:lit_compare_combined}
\begin{adjustbox}{width=1.4\textwidth}
\begin{threeparttable}
\small
\begin{tabular}{llllll}
\toprule
Study & Design / Sample & Paper estimate (SE) & Baseline mean & Percent effect & 95\% CI \\
\midrule

\multicolumn{6}{l}{\textit{Panel A. Participation}} \\
\midrule
This paper, 5 States & Admin data; DDD; age 45--49 vs.\ 50--55  & $-0.044$ $(0.005)$ & $0.667$ & $-6.6\%$ & [$-0.053$, $-0.034$]  \\
Gray et al.\ (Table 5), VA & Admin data; RD at age 50 & $-0.234$ $(0.015)$ & $0.632$ & $-37.0\%$ & [$-0.263$, $-0.205$] \\
Wheaton et al.\ (Table 4; Baseline from Table 3), AL & Admin data; Age 18--47 before reinstatement vs.\ after & $-0.28^{*}$ & 0.746 & $-37.5\%$ & n/r \\
Wheaton et al.\ (Table 4; Baseline from Table 3), CO G1 & Admin data; Age 18--47 before reinstatement vs.\ after & $0.0 $ & 0.469 & $0.0\%$ & n/r \\
Wheaton et al.\ (Table 4; Baseline from Table 3), CO G2 & Admin data; Age 18--47 before reinstatement vs.\ after & $-0.07^{*}$ & 0.492 & $-14.2\%$ & n/r \\
Wheaton et al.\ (Table 4; Baseline from Table 3), MD & Admin data; Age 18--47 before reinstatement vs.\ after & $-0.23^{*}$ & 0.682 & $-33.7\%$ & n/r \\
Wheaton et al.\ (Table 4; Baseline from Table 3), MN & Admin data; Age 18--47 before reinstatement vs.\ after & $-0.27^{*}$ & 0.670 & $-40.3\%$ & n/r \\
Wheaton et al.\ (Table 4; Baseline from Table 3), MO & Admin data; Age 18--47 before reinstatement vs.\ after & $-0.31^{*}$ & 0.715 & $-43.4\%$ & n/r \\
Wheaton et al.\ (Table 4; Baseline from Table 3), OR Narrow & Admin data; Age 18--47 before reinstatement vs.\ after & $-0.15 $ & 0.717 & $-20.9\%$ & n/r \\
Wheaton et al.\ (Table 4; Baseline from Table 3), OR Broad & Admin data; Age 18--47 before reinstatement vs.\ after & $-0.13^{*}$ & 0.741 & $-17.5\%$ & n/r \\
Wheaton et al.\ (Table 4; Baseline from Table 3), PA & Admin data; Age 18--47 before reinstatement vs.\ after & $-0.21^{*}$ & 0.665 & $-31.6\%$ & n/r \\
Wheaton et al.\ (Table 4; Baseline from Table 3), TN & Admin data; Age 18--47 before reinstatement vs.\ after & $-0.15^{*}$ & 0.730 & $-20.5\%$ & n/r \\
Wheaton et al.\ (Table 4; Baseline from Table 3), VT & Admin data; Age 18--47 before reinstatement vs.\ after & $-0.32^{*}$ & 0.714 & $-44.8\%$ & n/r \\
Han (Table 1), National & ACS; DDD; age 48--49 vs.\ 51--52 & $-0.015$ $(0.006)$ & $0.167$ & $-9.0\%$ & [$-0.027$, $-0.003$] \\
Stacy et al.\ (Table 3), 9 States & ACS+Admin; Diff-in-disc around age 50; bandwidth = 2 years & $-0.031$ $(0.018)$ & $0.201$ & $-15.4\%$ & [$-0.067$, $+0.004$] \\
Harris (Table A4), National & ACS; DD; age 25--49 & $-0.017$ $(0.004)$ & $0.183$ & $-9.5\%$ & [$-0.025$, $-0.009$] \\
Harris (Table A5), National & ACS; DDD; age 45--49 vs.\ 50--54 & $-0.015$ $(0.003)$ & $0.156$ & $-9.7\%$ & [$-0.021$, $-0.009$] \\
Cuffey et al.\ National & CPS; RD at age 50 & n/r & n/r & n/r & n/r \\
Ritter, Low-waiver state-year & CPS or SNAP QC; RD at age 50 & n/r & n/r & n/r & n/r \\
\midrule
\multicolumn{6}{l}{\textit{Panel B. Employment}} \\
\midrule
This paper, 5 States & Admin data; DDD; age 45--49 vs.\ 50--55  & $-0.004$ $(0.003)$ & $0.269$ & $-1.5\%$ & [$-0.010$, $0.002$] \\
Gray et al.\ (Table 5), VA & Admin data; RD at age 50 & $+0.010$ $(0.013)$ & $0.273$ & $+3.7\%$ & [$-0.015$, $+0.035$] \\
Wheaton et al.\ (Figure 28; Baseline from Table 3), CO G1 & Admin data; Age 18--47 before reinstatement vs.\ after & $-0.04^{*}$ & $0.366$ & $-10.9\%$ & n/r \\
Wheaton et al.\ (Figure 28; Baseline from Table 3), CO G2 & Admin data; Age 18--47 before reinstatement vs.\ after & $-0.06^{*}$ & $0.356$ & $-16.9\%$ & n/r \\
Wheaton et al.\ (Figure 28; Baseline from Table 3), MO & Admin data; Age 18--47 before reinstatement vs.\ after & $-0.03^{*}$ & $0.313$ & $-9.6\%$ & n/r \\
Wheaton et al.\ (Figure 28; Baseline from Table 3), PA & Admin data; Age 18--47 before reinstatement vs.\ after & $-0.02^{*}$ & $0.360$ & $-5.6\%$ & n/r \\
Han (Table 1), National & ACS; DDD; age 48--49 vs.\ 51--52 & $0.000$ $(0.007)$ & $0.747$ & $+0.0\%$ & [$-0.014$, $+0.014$] \\
Stacy et al.\ (Table 3), 9 States & ACS+Admin; Diff-in-disc around age 50; bandwidth = 2 years & $-0.015$ $(0.035)$ & $0.484$ & $-3.1\%$ & [$-0.084$, $+0.053$] \\
Harris (Table 2), National & ACS; DD; age 25--49 & $+0.013$ $(0.003)$ & $0.715$ & $+1.8\%$ & [$+0.007$, $+0.019$] \\
Harris (Table 3), National & ACS; DDD; age 45--49 vs.\ 50--54 & $+0.011$ $(0.005)$ & $0.725$ & $+1.5\%$ & [$+0.001$, $+0.021$] \\
Cuffey et al.\ (Table 3), National & CPS; RD at age 50 & $+0.130$ $(0.032)$ & $0.350$ & $+37.1\%$ & [$+0.067$, $+0.193$] \\
Ritter (Table 2; Baseline from Figure 3), Low-waiver state-year & CPS; RD at age 50; Men sample & $-0.021 (0.043)$ & $0.52$ & $-4.0\%$ & [$-0.105$, $+0.063$]  \\
Ritter (Table 3; Baseline from Figure 3), Low-waiver state-year & SNAP QC; RD at age 50; Men sample & $+0.007 (0.017)$ & $0.06$ & $+11.6\%$ & [$-0.026$, $+0.040$] \\ 
\bottomrule
\end{tabular}
\vspace{1.0em}
\begin{tablenotes}[flushleft]
\footnotesize
\item \textit{Notes:} ``n/r'' denotes not reported. Percent effects are computed as the estimate divided by the reported control mean or mean dependent variable. Estimates are coded as effects of imposing work requirements; signs are reversed when the original study reports effects of work-requirement exemptions or aging out of the requirement. Datasets: ACS = American Community Survey; CPS = Current Population Survey; SNAP QC = SNAP Quality Control data. Designs: DD = difference-in-differences; DDD = triple differences; RD = regression discontinuity; diff-in-disc = difference-in-discontinuities.
\item ${*}$: For Wheaton et al., standard errors are not reported; instead, we indicate with asterisk (${*}$) whether the effects are statistically significant at the 5\% level. 
\end{tablenotes}
\end{threeparttable}
\end{adjustbox}
\end{table}
\end{landscape}


\begin{table}[t]
\caption{Effects of SNAP Work Requirements on Contemporaneous Participant Characteristics}
\label{tab:composition_notfixed}
\vspace{-30bp}
\makebox[\textwidth][c]{\hspace*{1cm}\includegraphics[page=1,width=1.15\textwidth]{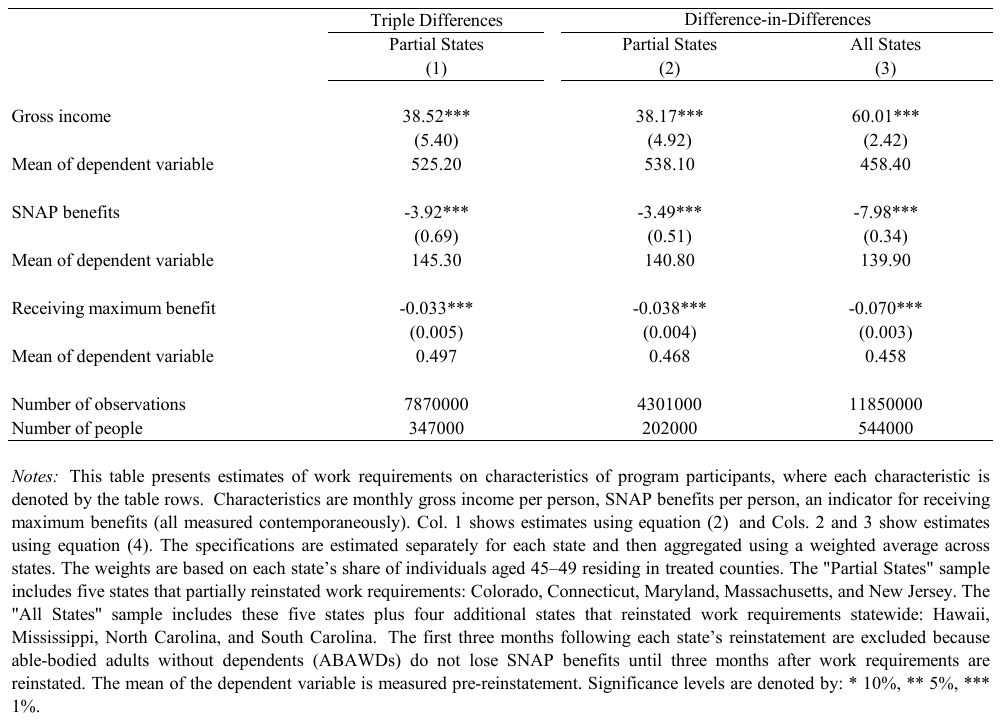}}
\begin{centering}
\begin{minipage}[t]{0.82\columnwidth}%
\end{minipage}
\par\end{centering}
\end{table}


\begin{table}[t]
\caption{Effects of SNAP Work Requirements on Participant Demographic Characteristics}
\label{tab:demographics}
\vspace{-60bp}
\includegraphics[page=1,width=1.1\textwidth]{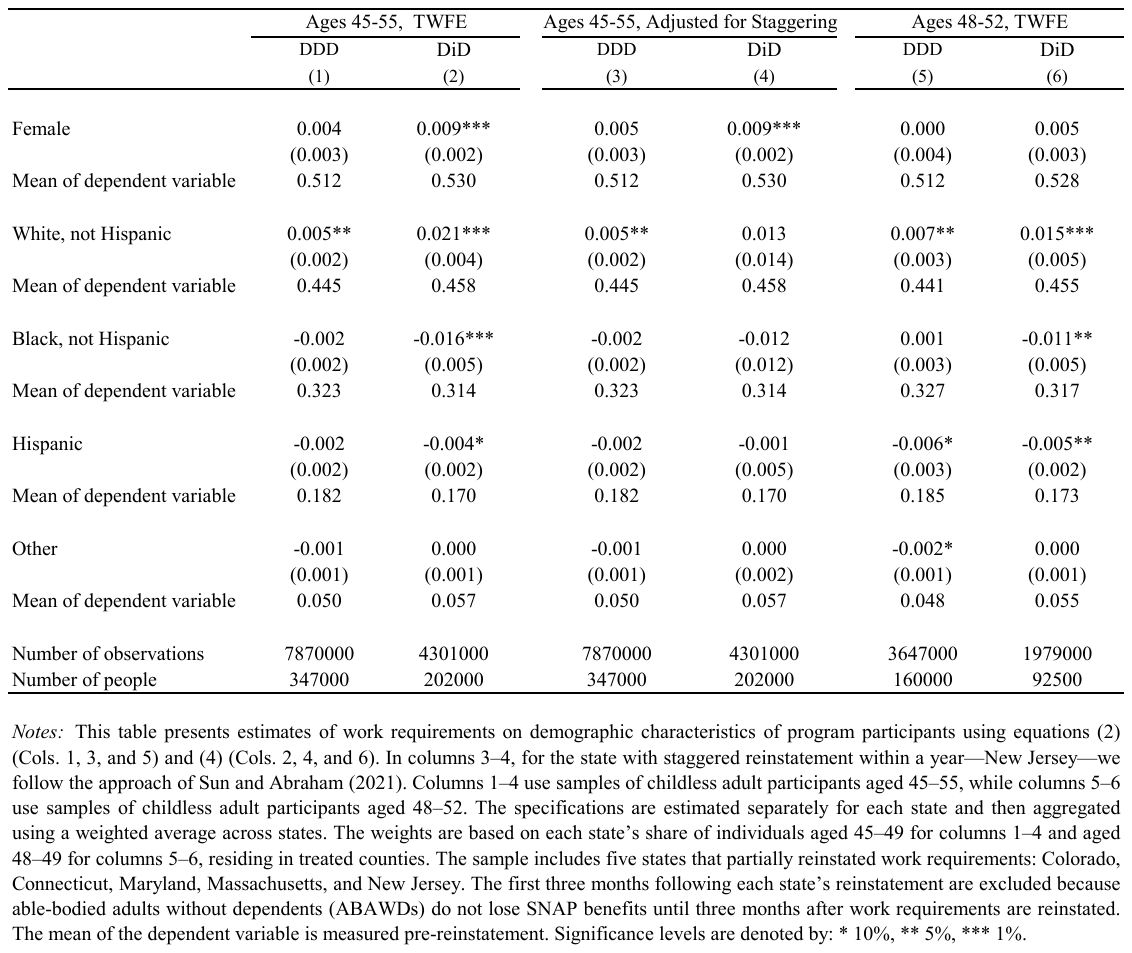}
\begin{centering}
\begin{minipage}[t]{0.82\columnwidth}%
\end{minipage}
\par\end{centering}
\end{table}


\begin{table}[t]
\caption{Estimated Effects of SNAP Work Requirements, Adjusted for Staggered Timing}
\label{tab:adj_staggering}
\vspace{-15bp}
\makebox[\textwidth][c]{\hspace*{2cm}\includegraphics[page=1,width=1.13\textwidth]{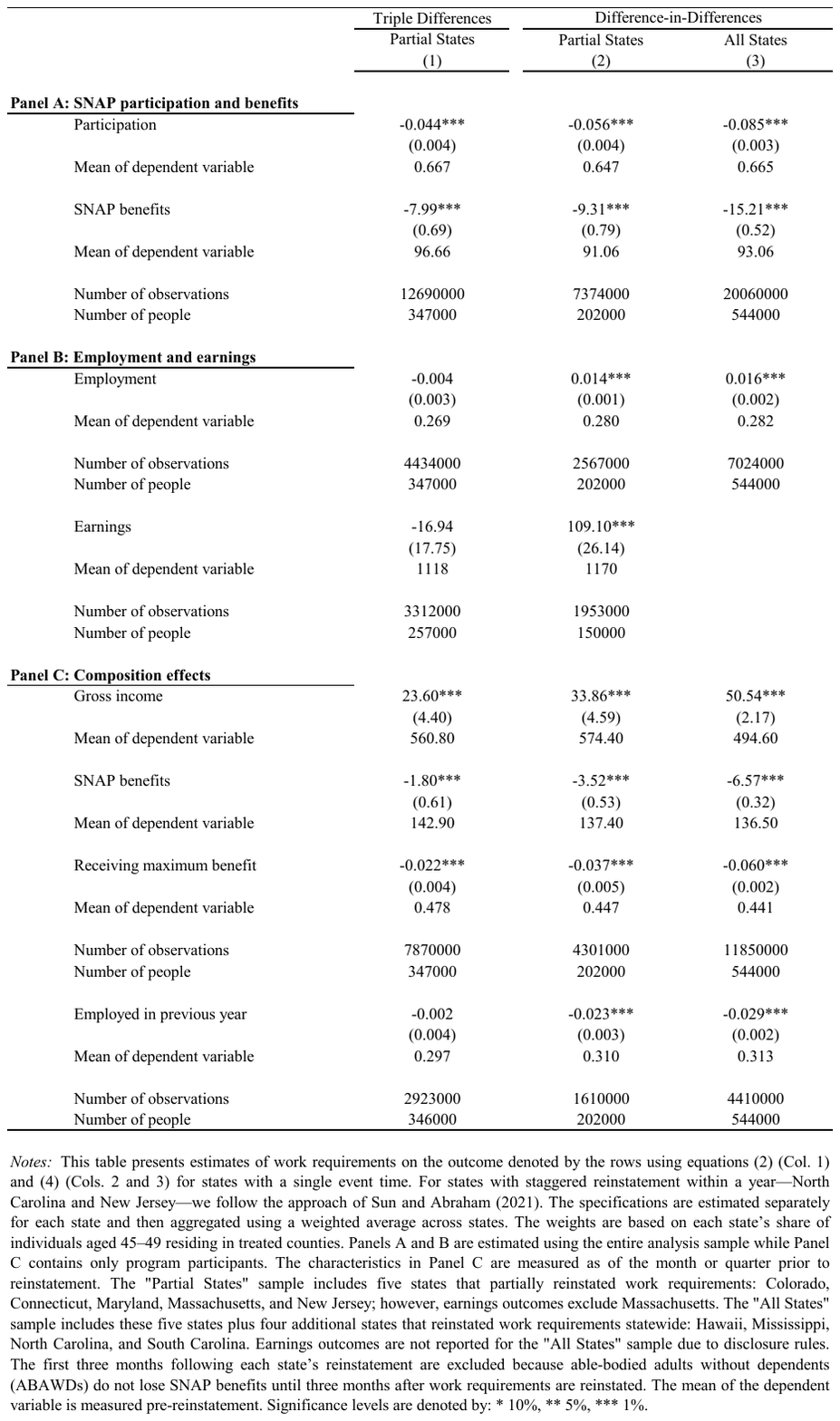}}
\begin{centering}
\begin{minipage}[t]{0.82\columnwidth}%
\end{minipage}
\par\end{centering}
\end{table}


\begin{table}[t]
\caption{Effects of SNAP Work Requirements, Ages 48 to 52}
\label{tab:ages48_to_52}
\vspace{-30bp}
\includegraphics[page=1,width=1.15\textwidth]{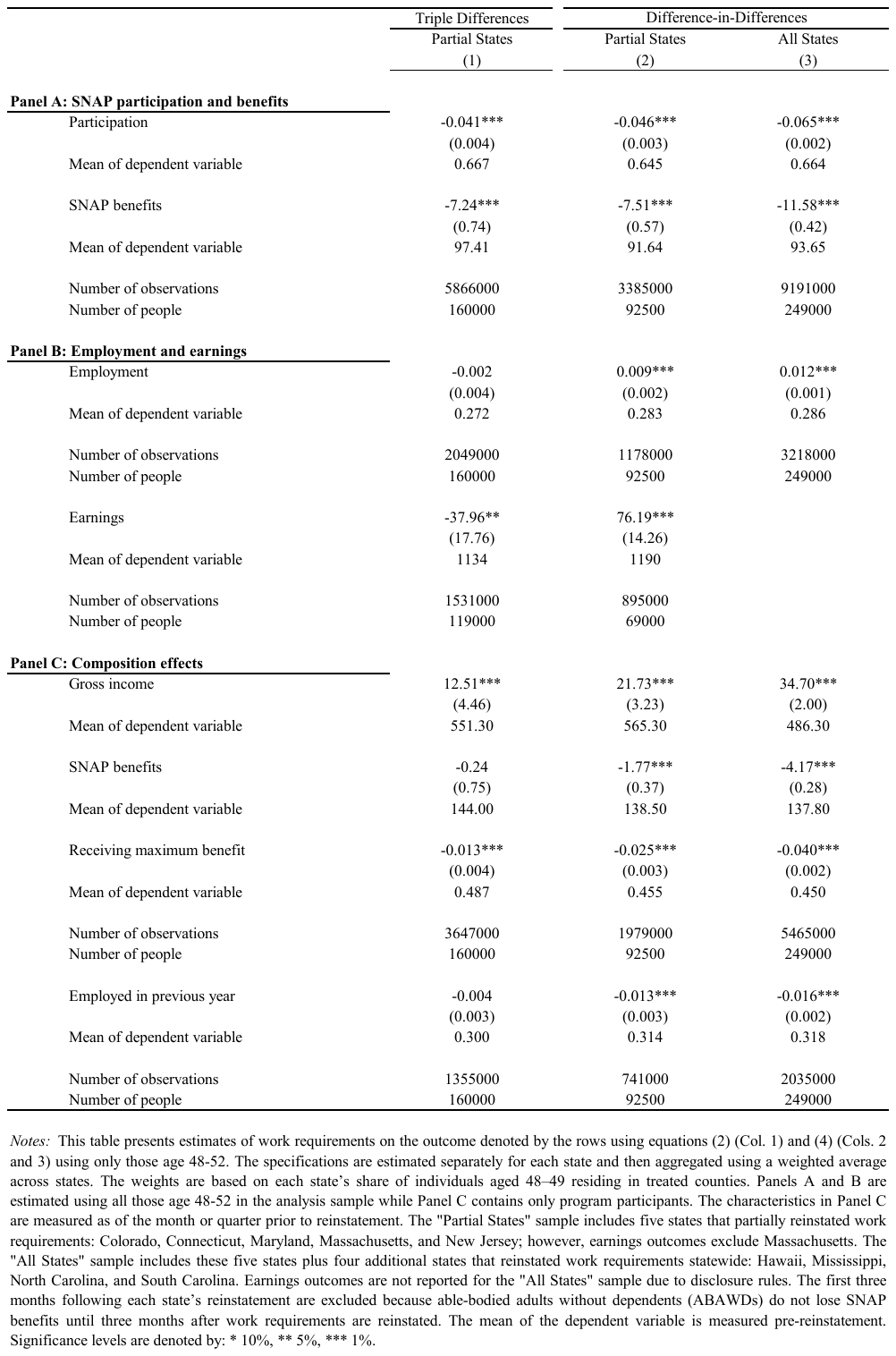}
\begin{centering}
\begin{minipage}[t]{0.82\columnwidth}%
\end{minipage}
\par\end{centering}
\end{table}



\counterwithin{figure}{section}

\end{document}